\documentclass[aps,prl,preprint,groupedaddress,linenumbers]{revtex4}

\usepackage{graphicx}
\usepackage{dcolumn}
\usepackage{bm}
\usepackage{amsmath}
\usepackage{color}
\usepackage{amssymb}
\usepackage{mathrsfs}
\usepackage{MnSymbol}
\usepackage{wasysym}
\usepackage{gensymb}
\usepackage{framed,color}
\usepackage{enumitem}
\usepackage{tabularx}
\usepackage{array}
\usepackage{float}
\usepackage{subcaption}
\usepackage{ulem}
\usepackage{hyperref}
\usepackage{setspace}
\usepackage{bbold}

\makeatletter
\newcommand{\colorcaption}[2][]{%
  \begingroup%
  \renewcommand{\@caption@fignum@sep}{ (color online). }%
  \caption[#1]{#2}%
  \endgroup%
  }
\makeatother

\makeatletter
\newcommand*{\rom}[1]{\expandafter\@slowromancap\romannumeral #1@}
\makeatother

\begin{document}


\title{Universal Scaling of the Little--Parks Oscillation Amplitude}

\author{Yuelin Shen,$^{1}$ Zixuan Huang,$^{1}$ Xiaoying Xu$^{2,\ddagger}$, Wei Qin,$^{3,\dagger}$ Yufan Li,$^{1,4,*}$}

\def\thefootnote{$\ddagger$}\footnotetext{Contact author: xuxiaoying@quantumsc.cn}\def\thefootnote{\arabic{footnote}}
\def\thefootnote{$\dagger$}\footnotetext{Contact author: qinwei5@ustc.edu.cn}\def\thefootnote{\arabic{footnote}}
\def\thefootnote{$*$}\footnotetext{Contact author: yufanli@cuhk.edu.hk}\def\thefootnote{\arabic{footnote}}

\affiliation{$^1$Department of Physics, The Chinese University of Hong Kong, Shatin, Hong Kong SAR, China}
\affiliation{$^2$Quantum Science Center of Guangdong-Hong Kong-Macao Greater Bay Area, Shenzhen, Guangdong, China}
\affiliation{$^3$International Center for Quantum Design of Functional Materials (ICQD), and Department of Physics, University of Science and Technology of China, Hefei, Anhui 230026, China}
\affiliation{$^4$State Key Laboratory of Quantum Information Technologies and Materials, Shatin, Hong Kong SAR, China}

\date{\today}

\begin{abstract}
    The Little--Parks effect played a pivotal role in establishing fluxoid quantization in doubly connected superconductors and providing early evidence for electron pairing, central to BCS theory.
    More recently, it has emerged as an important phase-sensitive probe of superconducting order. 
    Yet the amplitude of the Little--Parks oscillation has never been fully understood theoretically, as the conventional mean-field Ginzburg--Landau description does not account for its pronounced temperature dependence. 
    Here, we develop a universal theoretical framework incorporating thermal fluctuations in reduced-dimensional superconducting rings. 
    By analyzing Little--Parks measurements across multiple superconductors, we show that the resulting scaling quantitatively captures the temperature-dependent oscillation amplitudes over a broad range of materials and device geometries, resolving the long-standing discrepancy between theory and experiment.
\end{abstract}

\maketitle

\begin{text}

The Little--Parks effect is one of the earliest and most direct macroscopic manifestations of the quantum phase of the superconducting condensate. 
Dictated by the single-valuedness of the condensate wave function, the superfluid velocity of the Cooper pair condensate in a superconducting ring is periodic in the applied magnetic flux with period $\Phi_0=hc/2e$, leading to an oscillatory superconducting transition temperature $T_C$. 
In isothermal transport measurements, this oscillatory phase boundary appears as periodic magnetoresistance oscillations \cite{little_observation_1962}, providing early evidence for electron pairing and establishing the Little--Parks effect as a universal consequence of fluxoid quantization in doubly connected superconductors \cite{tinkham2004introduction}. 
Alongside its observation in a broad range of superconductors discovered over subsequent decades \cite{gammel_little-parks_1990,sochnikov_large_2010,carillo_little-parks_2010,Arpaia2014,cai_unconventional_2013,Yasui2017,qin_superconductivity_2017,Bi2024, Iwakiri2024}, the Little--Parks effect has attracted renewed interest as a versatile phase-sensitive probe of unconventional superconductivity. 
In purpose-designed superconducting rings, it can reveal sign changes in the superconducting gap function, fractional fluxoid states, and other unconventional phase structures that are inaccessible to conventional thermodynamic property measurements \cite{li_observation_2019,xu_spin-triplet_2020,Hua2023, Xu2024,li_unconventional_2024,xu_signature_2025,almoalem_observation_2024,wan_unconventional_2024,Tokuda2025,wang_switchable_2025}. 

Shortly after the original experiment by Little and Parks, a now-standard Ginzburg--Landau (GL) description of the Little--Parks effect was developed \cite{Tinkham1963,TINKHAM1964}.  
In this framework, the kinetic energy of the fluxoid-quantized supercurrent periodically suppresses the superconducting transition temperature, giving a maximum relative amplitude $\Delta T_C/T_C=\xi_0^2/4r^2$ for a thin-walled ring of radius $r$, with $\xi_0$ being the zero-temperature coherence length.  
This $(\xi_0/r)^2$ scaling is widely accepted as the conventional benchmark for interpreting the Little--Parks amplitude \cite{tinkham2004introduction}, yet its inconsistencies with experimental observations have remained conspicuous. 
Most notably, it predicts a temperature-independent $\Delta T_C$ oscillation amplitude, whereas a pronounced temperature dependence was reported already in the early Little--Parks experiments \cite{groff_fluxoid_1968}. 
The same scaling further implies that the Little--Parks effect should become exceedingly small in superconductors with short coherence lengths. 
High-$T_C$ cuprates provide a prime example, as their coherence lengths are typically only a few nanometers.
Indeed, when $\Phi_0$-periodic magnetoresistance oscillations were observed in La$_{2-x}$Sr$_x$CuO$_4$, they were attributed to vortex dynamics rather than to the Little--Parks effect itself, as their amplitude exceeded the conventional estimate by nearly two orders of magnitude \cite{sochnikov_large_2010,sochnikov_oscillatory_2010}.

Although conventional GL theory describes the thermodynamic phase boundary, Little--Parks experiments typically probe it by measuring resistance during the superconducting transition.
In reduced-dimensional superconducting structures, thermal fluctuations broaden this resistive transition, with thermally activated phase slips (TAPS) providing an important dissipation mechanism \cite{langer_Ambegaokar_1967,mccumber_intrinsic_1968}.
Consequently, the oscillation amplitude inferred from transport need not coincide with the mean-field shift of the thermodynamic transition temperature.
Here, we incorporate TAPS into the Ginzburg--Landau description and derive a universal scaling for the experimentally observed Little--Parks amplitude.
The resulting scaling accounts for the pronounced temperature dependence of the oscillation amplitude and quantitatively describes measurements across superconductors with widely different transition temperatures, coherence lengths, and device geometries. 
It thereby resolves the long-standing discrepancy between conventional theory and experiment and unifies $\Phi_0$-periodic magnetoresistance oscillations observed in superconducting rings within a universal framework for the Little--Parks effect. 

The $T_C$ oscillation of a thin-walled superconducting ring near the phase transition is conventionally described by the linearized Ginzburg--Landau (GL) theory \cite{moshchalkov_effect_1995},
\begin{equation}\label{LP}
\frac{T_C(H)-T_C(0)}{T_C(0)}
=
\frac{\xi_0^2}{r^2}
\left[
-\left(\frac{w}{2r}\right)^2
\left(\frac{\Phi}{\Phi_0}\right)^2
-
\left(n-\frac{\Phi}{\Phi_0}\right)^2
\right],
\end{equation}
where $\Phi$ is the applied magnetic flux threading the ring; $r$ and $w$ are the ring radius and width, respectively; $n$ is an integer fluxoid number; and $\xi_0$ is the zero-temperature coherence length. 
The first term describes the non-oscillatory background of the Little--Parks effect. 
The second term gives the Little--Parks oscillation, with a period of $\Phi_0$ and a maximum transition-temperature amplitude $\Delta T_C=\frac{\xi_0^2}{4r^2}T_C$.
The factor $\frac{1}{4}$ comes from the maximum value of $(n-\frac{\Phi}{\Phi_0})^2$. 
For a given superconducting material, $\xi_0$ and $T_C$ are fixed, while a specified ring geometry fixes $r$; Eq.~\eqref{LP} therefore predicts a uniquely determined oscillation amplitude.
In this conventional description, neither the relative amplitude $\Delta T_C/T_C$ nor the absolute amplitude $\Delta T_C$ contains any dependence on the measurement temperature.
In transport measurements, the Little--Parks effect is observed within a finite temperature window along the $R$--$T$ transition, where superconducting order coexists with a finite resistance.
To see how Eq.~\eqref{LP} holds up against temperature evolution, we present in Figs.~1(a) and 1(b) the Little--Parks oscillations measured at several temperatures in Nb and $\beta$-Bi$_2$Pd rings, respectively.
Nb is a conventional superconductor with $T_C \simeq$7.5~K, whereas $\beta$-Bi$_2$Pd is an unconventional superconductor \cite{li_observation_2019,Xu2024} with $T_C \simeq$3.5~K.
For each device, the amplitude of the $\Phi_0$-periodic oscillations varies drastically with temperature. 
In the Nb ring, $\Delta T_C$ decreases by a factor of 4.3 over a temperature increase of only 0.11~K.
A similar trend is observed in the $\beta$-Bi$_2$Pd ring, where $\Delta T_C$ is suppressed by a factor of 3.6 within a temperature window of 0.20~K.  
Thus, in both systems the Little--Parks amplitude is strongly temperature dependent, in direct contrast to Eq.~\eqref{LP}.
Taking $\xi_0\simeq7.0$~nm for Nb \cite{supple} and $\xi_0\simeq13.7$~nm for $\beta$-Bi$_2$Pd \cite{Powell2025}, 
Eq.~\eqref{LP} predicts temperature-independent amplitudes of $\Delta T_C=8.27\times10^{-4}$~K and $2.46\times10^{-3}$~K for the Nb and $\beta$-Bi$_2$Pd rings, respectively.
These GL values are also substantially smaller than the measured oscillation amplitudes indicated in Fig.~1.

The inadequacy of the conventional Little--Parks amplitude description is not limited to these two systems.
We survey the reported experimental data available in the literature and analyze those that are (i) obtained from devices with well-defined ring geometries; (ii) measured at multiple temperatures; and (iii) reported as magnetoresistance together with the corresponding $R$--$T$ transition data.
Fig.~1(c) summarizes the Little--Parks amplitude as a function of reduced temperature for Sn \cite{groff_fluxoid_1968,parks_fluxoid_1964}, $\alpha$-BiPd \cite{xu_spin-triplet_2020}, $\beta$-Bi$_2$Pd, Nb, La$_{2-x}$Sr$_x$CuO$_4$ (LSCO) \cite{sochnikov_large_2010,sochnikov_oscillatory_2010} and YBa$_2$Cu$_3$O$_{7-\delta}$ (YBCO) \cite{gammel_little-parks_1990,Arpaia2014}.
These materials encompass both conventional and unconventional superconductors, spanning low-$T_C$ and high-$T_C$ systems.
Across all of the systems surveyed, two common features emerge: a pronounced temperature dependence of the oscillation amplitude and a substantial departure in magnitude from the temperature-independent GL expectation.
Taken together, these observations point to a common missing ingredient in the conventional description of the Little--Parks amplitude.

The failure of Eq.~\eqref{LP} to describe the measured Little--Parks amplitude indicates that the thermodynamic $T_C(\Phi)$ alone is insufficient to describe the transport experiment.
We therefore turn directly to the flux dependence of the resistance $R(T,\Phi)$.
Near the superconducting transition, thermally activated phase slips (TAPS) provide a natural mechanism for finite resistance in reduced-dimensional superconducting structures \cite{langer_Ambegaokar_1967,mccumber_intrinsic_1968,tinkham2004introduction,lukens_onset_1970}.
The fluxoid-quantized supercurrent modifies the free-energy landscape of a phase slip and hence its activation barrier; because the supercurrent varies periodically with $\Phi$, so does the resulting resistance.

Assuming that the thermally activated phase slip gives the dominant contribution to the resistance in the Little--Parks experiments,  
we first consider the temperature dependence of the resistance below the thermodynamic critical temperature.
For a quasi-one dimensional system, the temperature dependence of the resistance below the thermodynamic critical temperature is given by
\begin{equation}
R(T)
=
R_n\exp\left
(-\frac{\Delta F_0(T)}{K_B T}
\right)
,
\label{RT}
\end{equation}
where $R_n$ denotes the normal-state resistance and $\Delta F_0 (T)$ represents the phase-slip fluctuation energy barrier encountered when transitioning through an intermediate state. The superconducting order parameter is suppressed locally and momentarily in this intermediate state. The free energy of the intermediate state constitutes the energy barrier given by $\Delta F_0 (T)=\sqrt{2} A\xi H_C^2/3\pi$, where $A$ denotes the cross-sectional area of the ring, $H_C$ is the critical magnetic field, and the coherence length $\xi=\xi_0/\sqrt{(1-t)}$ with the reduced temperature $t=T/T_{C0}$. 
The presence of magnetic flux $\Phi$ drives the supercurrent to circulate the ring, leading to an overall enhancement of the GL free energy $\Delta F_{\mathrm{kin}}(T,\Phi)$. 
This enhancement will essentially reduce the phase-slip free energy barrier, giving rise to an enhancement of resistance
\begin{equation}
R(T,\Phi)
=
R_n \exp\left
(-\frac{(\Delta F_0 (T)-\Delta F_{\mathrm{kin}}(T,\Phi)
}{k_B T}
\right)
.
\label{RTphi}
\end{equation}
Since $\Delta F_{\mathrm{kin}}(T,\Phi)$ oscillates with $\Phi$, the experimentally measured resistance $R(T,\Phi)$ exhibits an oscillatory behavior, with a period identical to the oscillation of $T_C$ predicted by GL theory.

Having established the temperature- and flux-dependent resistance, we now determine the Little--Parks amplitude as it is extracted experimentally.
For a given temperature $T$, the experimentally identified $\Delta T_C$ is obtained by comparing the resistances with and without magnetic flux, $R(T+\Delta T_C,\Phi)=R(T,0)$. 
By expanding $R(T+\Delta T_C ,\Phi)$ around $T$ with respect to $\Delta T_C$, we finally have the Little--Parks oscillation of $\Delta T_C$. 
Therefore, for a conventional superconducting ring, we have
\begin{equation}
\frac{\Delta T_C}{T_C}
=
\chi (t)\frac{\xi_0}{r}
\left[
-\left(\frac{w}{2r}\right)^2
\left(\frac{\Phi}{\Phi_0}\right)^2
-
\left(n-\frac{\Phi}{\Phi_0}\right)^2
\right]
,
\label{newLP}
\end{equation}
with $\chi(t)=(3\pi /\sqrt2)\cdot (t\sqrt{1-t}/(t+2))$.
The detailed derivation is provided in the Supplemental Material \cite{supple}.
In this expression, the flux--dependent part retains the same form as in the conventional Little--Parks theory, while the oscillation amplitude is modified to $\Delta T_C=\chi (t)\frac{\xi_0}{4r}T_C$. 
Here, the factor $\frac{1}{4}$ arises from the maximum value of $\left(n-\frac{\Phi}{\Phi_0}\right)^2$, which reaches $\frac{1}{4}$ halfway between adjacent fluxoid states.
The resulting oscillation amplitude is therefore temperature-dependent through the fluctuation prefactor $\chi(t)$.

We examine the experimental data against Eq.~\eqref{newLP}. 
Figs.~2(a) and 2(b) present the temperature evolution of the Little--Parks oscillation amplitude for the Nb and $\beta$-Bi$_2$Pd rings, respectively. 
In both devices, the oscillation amplitude $\Delta T_C$ decreases with increasing temperature. 
Following Eq.~\eqref{newLP}, we fit the temperature-dependent amplitudes using 
\begin{equation}
\Delta T_C
=
\chi (t)\frac{\xi_0}{4r}T_{C0}, 
\label{Amplitude}
\end{equation}
with the coherence length $\xi_0$ and the zero-field transition temperature $T_{c0}$ of the ring structure being the fitting parameters.
For Nb, the fit yields $\xi_0=7.3$~nm, in excellent agreement with $\xi_0=7.0$~nm obtained independently from the $H_{C2}$ measurement.
The fitted $T_{C0}$ is expected to be lower than the transition temperature determined from the $R$--$T$ curve of the entire device, since the latter also contains contributions from the wider contact pads, which generally become superconducting at higher temperatures than the narrow ring. 
Using the fitted values $\xi_0 =$7.3~nm and $T_{c0} =$ 6.08~K, Eq.~\eqref{newLP} further reproduces the Little--Parks oscillations measured at different temperatures, as shown in Fig.~2(c). 
The experimental data can be well accounted for across all temperatures. 
For $\beta$-Bi$_2$Pd, fitting with Eq.~\eqref{Amplitude} yields $\xi_0$=19.0~nm, as shown in Fig.~2(b), consistent with the value ($\xi_0$=13.7~nm) obtained from the $H_{C2}$ measurements. 
Using this fitted value, the Little--Parks oscillations at different temperatures are likewise well reproduced, as shown in Fig.~2(d). 
For $\beta$-Bi$_2$Pd, we additionally account for the linear-in-field background modulation reported in our previous work \cite{xu_signature_2025}, which is absent from the conventional Little--Parks expression.
The corresponding modified form of Eq.~\eqref{newLP} is given in the Supplemental Material \cite{supple}.

We apply the same analysis to the Little--Parks data collected from the literature.
For each dataset, the temperature-dependent amplitude is first fitted using Eq.~\eqref{Amplitude}, and the extracted $\xi_0$ is checked against independently reported values where available. The same parameters are then used in Eq.~\eqref{newLP} to reproduce the flux-dependent oscillations.
Across the materials surveyed, the measured oscillations are well described over most of the accessible temperature range.
The corresponding fits are summarized in the Supplemental Material \cite{supple}.

The universality of the new scaling is most clearly revealed by normalizing the Little--Parks amplitude as $\frac{\Delta T_C}{T_C}\frac{4r}{\xi_0}$, which removes the material- and geometry-dependent factor $\xi_0/r$.
According to Eq.~\eqref{newLP}, all data should then follow the same reduced-temperature dependence $\chi(t)$.
In Fig.~3(a) we summarize the normalized amplitudes as a function of $T/T_C$. 
The result demonstrates remarkable consistency as all data collapse around a common temperature-dependent curve well described by $\chi(t)$; despite the fact that there are more than one order of magnitude variations in $\xi_0$ as well as in ring dimensions. 
This collapse demonstrates that the temperature dependence is captured by the universal function $\chi(t)$, while the material and geometric dependence is governed by the single scaling variable $\xi_0/r$.

For comparison, we normalize the same data according to the conventional GL expression, $\frac{\Delta T_C}{T_C}\frac{4r^2}{\xi_0^2}$.
The result is shown in Fig.~3(b). 
Equation~\eqref{LP} predicts that the normalized amplitude should be temperature independent and equal to unity, as indicated by the gray dashed line.
Instead, the data retain a pronounced temperature dependence and scatter strongly among different materials and ring sizes.
Thus, the conventional description fails in two respects: it does not capture the temperature dependence of the measured amplitude, and the scaling factor $(\xi_0/r)^2$ does not account for its material and geometric dependence.

The contrast between the new scaling factor $\chi (t)\frac{\xi_0}{r}$ and the conventional $\frac{\xi_0^2}{r^2}$ becomes most pronounced when $\frac{\xi_0}{r} \ll 1$. 
Mesoscopic superconducting rings typically have radii of order $10^2$--$10^3$~nm, so the distinction between the two scalings is expected to be greatest in superconductors with the shortest coherence lengths, most notably the high-$T_C$ cuprates.
Consistent with this expectation, Fig.~3(b) shows the largest deviations from the conventional GL prediction in YBCO and LSCO, reaching nearly two orders of magnitude for larger rings.
The difficulty of reconciling such large oscillation amplitudes with the conventional Little--Parks estimate was recognized previously and motivated an alternative interpretation based on vortex dynamics \cite{sochnikov_large_2010,sochnikov_oscillatory_2010}.
In that picture, the magnetoresistance oscillations in LSCO arise from a periodic modulation of the activation barrier for thermally excited vortex motion by the fluxoid-quantized persistent current, rather than from an oscillation of $\Delta T_C$.
The universal scaling established in Fig.~3(a), however, leads to a different interpretation.
Once thermal fluctuations in the resistive transition are taken into account, the Little--Parks amplitude scales as $\chi(t)\xi_0/r$ rather than $(\xi_0/r)^2$, substantially reducing the suppression expected for short-coherence-length superconductors.
The LSCO and YBCO results then fall naturally within the same framework as low-$T_C$ superconductors, without requiring a distinct origin for their $\Phi_0$-periodic magnetoresistance oscillations.
Using the universal scaling, the fitted coherence lengths are $\xi_0\approx1.1$~nm for LSCO and $\xi_0\approx3.2$~nm for YBCO, both consistent with the short, nanometer-scale coherence lengths reported for high-$T_C$ cuprates.

Despite the much higher transition temperatures of LSCO and YBCO, most of the Little--Parks oscillations considered here are observed within a similar reduced-temperature window, approximately $0.85<T/T_C<1$, as shown in Fig.~3.
This corresponds to the finite-resistance regime in which thermally activated phase slips are expected to provide an important contribution to the transport response.
At the high-temperature side of the resistive transition, where the resistance approaches its normal-state value, the description in terms of thermally activated transitions between well-defined superconducting states becomes less applicable.
At sufficiently low temperatures, on the other hand, thermal activation is progressively suppressed and other dissipation mechanisms, including quantum phase slips in sufficiently narrow structures, may become important \cite{giordano_evidence_1988,arutyunov_superconductivity_2008}.
The experimentally observed Little--Parks oscillations therefore predominantly fall within the intermediate transition regime in which the present TAPS-based scaling is expected to apply.

In summary, we show that the conventional mean-field GL description fails to capture the Little--Parks oscillation amplitude, both in its magnitude and its pronounced temperature dependence.
By incorporating TAPS into the description of the resistive transition, we establish a universal scaling of the amplitude governed by $\chi(t)\xi_0/r$.
This scaling accounts for both the strong temperature dependence observed experimentally and the large departures from the conventional $(\xi_0/r)^2$ expectation. 
The same scaling quantitatively describes superconductors spanning widely different transition temperatures, coherence lengths, device dimensions, and pairing symmetries, including the high-$T_C$ cuprates for which the conventional estimate fails most dramatically.
These results restore the $\Phi_0$-periodic magnetoresistance oscillations observed across these systems to a common Little--Parks framework and provide a quantitative basis for interpreting future phase-sensitive measurements in mesoscopic superconducting devices.

\end{text}

\bibliography{amplitude_LP}

\begin{thebibliography}{34}
\expandafter\ifx\csname natexlab\endcsname\relax\def\natexlab#1{#1}\fi
\expandafter\ifx\csname bibnamefont\endcsname\relax
  \def\bibnamefont#1{#1}\fi
\expandafter\ifx\csname bibfnamefont\endcsname\relax
  \def\bibfnamefont#1{#1}\fi
\expandafter\ifx\csname citenamefont\endcsname\relax
  \def\citenamefont#1{#1}\fi
\expandafter\ifx\csname url\endcsname\relax
  \def\url#1{\texttt{#1}}\fi
\expandafter\ifx\csname urlprefix\endcsname\relax\def\urlprefix{URL }\fi
\providecommand{\bibinfo}[2]{#2}
\providecommand{\eprint}[2][]{\url{#2}}

\bibitem[{\citenamefont{Little and Parks}(1962)}]{little_observation_1962}
\bibinfo{author}{\bibfnamefont{W.~A.} \bibnamefont{Little}} \bibnamefont{and}
  \bibinfo{author}{\bibfnamefont{R.~D.} \bibnamefont{Parks}},
  \bibinfo{journal}{Physical Review Letters} \textbf{\bibinfo{volume}{9}},
  \bibinfo{pages}{9} (\bibinfo{year}{1962}),
  \urlprefix\url{https://link.aps.org/doi/10.1103/PhysRevLett.9.9}.

\bibitem[{\citenamefont{Tinkham}(2004)}]{tinkham2004introduction}
\bibinfo{author}{\bibfnamefont{M.}~\bibnamefont{Tinkham}},
  \emph{\bibinfo{title}{Introduction to Superconductivity}}
  (\bibinfo{publisher}{Dover Publications}, \bibinfo{year}{2004}),
  \bibinfo{edition}{2nd} ed., ISBN \bibinfo{isbn}{0486435032},
  \urlprefix\url{http://www.worldcat.org/isbn/0486435032}.

\bibitem[{\citenamefont{Gammel et~al.}(1990)\citenamefont{Gammel, Polakos,
  Rice, Harriott, and Bishop}}]{gammel_little-parks_1990}
\bibinfo{author}{\bibfnamefont{P.~L.} \bibnamefont{Gammel}},
  \bibinfo{author}{\bibfnamefont{P.~A.} \bibnamefont{Polakos}},
  \bibinfo{author}{\bibfnamefont{C.~E.} \bibnamefont{Rice}},
  \bibinfo{author}{\bibfnamefont{L.~R.} \bibnamefont{Harriott}},
  \bibnamefont{and} \bibinfo{author}{\bibfnamefont{D.~J.}
  \bibnamefont{Bishop}}, \bibinfo{journal}{Physical Review B}
  \textbf{\bibinfo{volume}{41}}, \bibinfo{pages}{2593} (\bibinfo{year}{1990}),
  ISSN \bibinfo{issn}{0163-1829, 1095-3795},
  \urlprefix\url{https://link.aps.org/doi/10.1103/PhysRevB.41.2593}.

\bibitem[{\citenamefont{Sochnikov
  et~al.}(2010{\natexlab{a}})\citenamefont{Sochnikov, Shaulov, Yeshurun,
  Logvenov, and Božović}}]{sochnikov_large_2010}
\bibinfo{author}{\bibfnamefont{I.}~\bibnamefont{Sochnikov}},
  \bibinfo{author}{\bibfnamefont{A.}~\bibnamefont{Shaulov}},
  \bibinfo{author}{\bibfnamefont{Y.}~\bibnamefont{Yeshurun}},
  \bibinfo{author}{\bibfnamefont{G.}~\bibnamefont{Logvenov}}, \bibnamefont{and}
  \bibinfo{author}{\bibfnamefont{I.}~\bibnamefont{Božović}},
  \bibinfo{journal}{Nature Nanotechnology} \textbf{\bibinfo{volume}{5}},
  \bibinfo{pages}{516} (\bibinfo{year}{2010}{\natexlab{a}}), ISSN
  \bibinfo{issn}{1748-3395},
  \urlprefix\url{https://www.nature.com/articles/nnano.2010.111}.

\bibitem[{\citenamefont{Carillo et~al.}(2010)\citenamefont{Carillo, Papari,
  Stornaiuolo, Born, Montemurro, Pingue, Beltram, and
  Tafuri}}]{carillo_little-parks_2010}
\bibinfo{author}{\bibfnamefont{F.}~\bibnamefont{Carillo}},
  \bibinfo{author}{\bibfnamefont{G.}~\bibnamefont{Papari}},
  \bibinfo{author}{\bibfnamefont{D.}~\bibnamefont{Stornaiuolo}},
  \bibinfo{author}{\bibfnamefont{D.}~\bibnamefont{Born}},
  \bibinfo{author}{\bibfnamefont{D.}~\bibnamefont{Montemurro}},
  \bibinfo{author}{\bibfnamefont{P.}~\bibnamefont{Pingue}},
  \bibinfo{author}{\bibfnamefont{F.}~\bibnamefont{Beltram}}, \bibnamefont{and}
  \bibinfo{author}{\bibfnamefont{F.}~\bibnamefont{Tafuri}},
  \bibinfo{journal}{Physical Review B} \textbf{\bibinfo{volume}{81}},
  \bibinfo{pages}{054505} (\bibinfo{year}{2010}), ISSN
  \bibinfo{issn}{1098-0121, 1550-235X},
  \urlprefix\url{https://link.aps.org/doi/10.1103/PhysRevB.81.054505}.

\bibitem[{\citenamefont{Arpaia et~al.}(2014)\citenamefont{Arpaia, Charpentier,
  Toskovic, Bauch, and Lombardi}}]{Arpaia2014}
\bibinfo{author}{\bibfnamefont{R.}~\bibnamefont{Arpaia}},
  \bibinfo{author}{\bibfnamefont{S.}~\bibnamefont{Charpentier}},
  \bibinfo{author}{\bibfnamefont{R.}~\bibnamefont{Toskovic}},
  \bibinfo{author}{\bibfnamefont{T.}~\bibnamefont{Bauch}}, \bibnamefont{and}
  \bibinfo{author}{\bibfnamefont{F.}~\bibnamefont{Lombardi}},
  \bibinfo{journal}{Physica C: Superconductivity and its Applications}
  \textbf{\bibinfo{volume}{506}}, \bibinfo{pages}{184} (\bibinfo{year}{2014}),
  ISSN \bibinfo{issn}{0921-4534}.

\bibitem[{\citenamefont{Cai et~al.}(2013)\citenamefont{Cai, Ying, Staley, Xin,
  Fobes, Liu, Mao, and Liu}}]{cai_unconventional_2013}
\bibinfo{author}{\bibfnamefont{X.}~\bibnamefont{Cai}},
  \bibinfo{author}{\bibfnamefont{Y.~A.} \bibnamefont{Ying}},
  \bibinfo{author}{\bibfnamefont{N.~E.} \bibnamefont{Staley}},
  \bibinfo{author}{\bibfnamefont{Y.}~\bibnamefont{Xin}},
  \bibinfo{author}{\bibfnamefont{D.}~\bibnamefont{Fobes}},
  \bibinfo{author}{\bibfnamefont{T.~J.} \bibnamefont{Liu}},
  \bibinfo{author}{\bibfnamefont{Z.~Q.} \bibnamefont{Mao}}, \bibnamefont{and}
  \bibinfo{author}{\bibfnamefont{Y.}~\bibnamefont{Liu}},
  \bibinfo{journal}{Physical Review B} \textbf{\bibinfo{volume}{87}},
  \bibinfo{pages}{081104} (\bibinfo{year}{2013}), ISSN
  \bibinfo{issn}{1098-0121, 1550-235X},
  \urlprefix\url{https://link.aps.org/doi/10.1103/PhysRevB.87.081104}.

\bibitem[{\citenamefont{Yasui et~al.}(2017)\citenamefont{Yasui, Lahabi, Anwar,
  Nakamura, Yonezawa, Terashima, Aarts, and Maeno}}]{Yasui2017}
\bibinfo{author}{\bibfnamefont{Y.}~\bibnamefont{Yasui}},
  \bibinfo{author}{\bibfnamefont{K.}~\bibnamefont{Lahabi}},
  \bibinfo{author}{\bibfnamefont{M.~S.} \bibnamefont{Anwar}},
  \bibinfo{author}{\bibfnamefont{Y.}~\bibnamefont{Nakamura}},
  \bibinfo{author}{\bibfnamefont{S.}~\bibnamefont{Yonezawa}},
  \bibinfo{author}{\bibfnamefont{T.}~\bibnamefont{Terashima}},
  \bibinfo{author}{\bibfnamefont{J.}~\bibnamefont{Aarts}}, \bibnamefont{and}
  \bibinfo{author}{\bibfnamefont{Y.}~\bibnamefont{Maeno}},
  \bibinfo{journal}{Physical Review B} \textbf{\bibinfo{volume}{96}},
  \bibinfo{pages}{180507} (\bibinfo{year}{2017}), ISSN
  \bibinfo{issn}{2469-9969}.

\bibitem[{\citenamefont{Qin et~al.}(2017)\citenamefont{Qin, Shi, Ideue,
  Yoshida, Zak, Tenne, Kikitsu, Inoue, Hashizume, and
  Iwasa}}]{qin_superconductivity_2017}
\bibinfo{author}{\bibfnamefont{F.}~\bibnamefont{Qin}},
  \bibinfo{author}{\bibfnamefont{W.}~\bibnamefont{Shi}},
  \bibinfo{author}{\bibfnamefont{T.}~\bibnamefont{Ideue}},
  \bibinfo{author}{\bibfnamefont{M.}~\bibnamefont{Yoshida}},
  \bibinfo{author}{\bibfnamefont{A.}~\bibnamefont{Zak}},
  \bibinfo{author}{\bibfnamefont{R.}~\bibnamefont{Tenne}},
  \bibinfo{author}{\bibfnamefont{T.}~\bibnamefont{Kikitsu}},
  \bibinfo{author}{\bibfnamefont{D.}~\bibnamefont{Inoue}},
  \bibinfo{author}{\bibfnamefont{D.}~\bibnamefont{Hashizume}},
  \bibnamefont{and} \bibinfo{author}{\bibfnamefont{Y.}~\bibnamefont{Iwasa}},
  \bibinfo{journal}{Nature Communications} \textbf{\bibinfo{volume}{8}},
  \bibinfo{pages}{14465} (\bibinfo{year}{2017}), ISSN
  \bibinfo{issn}{2041-1723}, \bibinfo{note}{number: 1},
  \urlprefix\url{https://www.nature.com/articles/ncomms14465}.

\bibitem[{\citenamefont{Bi et~al.}(2024)\citenamefont{Bi, Zhang, Ao, Li, Huang,
  Qin, and Yuan}}]{Bi2024}
\bibinfo{author}{\bibfnamefont{X.}~\bibnamefont{Bi}},
  \bibinfo{author}{\bibfnamefont{Y.}~\bibnamefont{Zhang}},
  \bibinfo{author}{\bibfnamefont{L.}~\bibnamefont{Ao}},
  \bibinfo{author}{\bibfnamefont{H.}~\bibnamefont{Li}},
  \bibinfo{author}{\bibfnamefont{J.}~\bibnamefont{Huang}},
  \bibinfo{author}{\bibfnamefont{F.}~\bibnamefont{Qin}}, \bibnamefont{and}
  \bibinfo{author}{\bibfnamefont{H.}~\bibnamefont{Yuan}},
  \bibinfo{journal}{Advanced Functional Materials}
  \textbf{\bibinfo{volume}{35}} (\bibinfo{year}{2024}), ISSN
  \bibinfo{issn}{1616-3028}.

\bibitem[{\citenamefont{Iwakiri et~al.}(2024)\citenamefont{Iwakiri,
  Mestre-Torà, Portolés, Visscher, Perego, Zheng, Taniguchi, Watanabe,
  Sigrist, Ihn et~al.}}]{Iwakiri2024}
\bibinfo{author}{\bibfnamefont{S.}~\bibnamefont{Iwakiri}},
  \bibinfo{author}{\bibfnamefont{A.}~\bibnamefont{Mestre-Torà}},
  \bibinfo{author}{\bibfnamefont{E.}~\bibnamefont{Portolés}},
  \bibinfo{author}{\bibfnamefont{M.}~\bibnamefont{Visscher}},
  \bibinfo{author}{\bibfnamefont{M.}~\bibnamefont{Perego}},
  \bibinfo{author}{\bibfnamefont{G.}~\bibnamefont{Zheng}},
  \bibinfo{author}{\bibfnamefont{T.}~\bibnamefont{Taniguchi}},
  \bibinfo{author}{\bibfnamefont{K.}~\bibnamefont{Watanabe}},
  \bibinfo{author}{\bibfnamefont{M.}~\bibnamefont{Sigrist}},
  \bibinfo{author}{\bibfnamefont{T.}~\bibnamefont{Ihn}}, \bibnamefont{et~al.},
  \bibinfo{journal}{Nature Communications} \textbf{\bibinfo{volume}{15}}
  (\bibinfo{year}{2024}), ISSN \bibinfo{issn}{2041-1723}.

\bibitem[{\citenamefont{Li et~al.}(2019)\citenamefont{Li, Xu, Lee, Chu, and
  Chien}}]{li_observation_2019}
\bibinfo{author}{\bibfnamefont{Y.}~\bibnamefont{Li}},
  \bibinfo{author}{\bibfnamefont{X.}~\bibnamefont{Xu}},
  \bibinfo{author}{\bibfnamefont{M.-H.} \bibnamefont{Lee}},
  \bibinfo{author}{\bibfnamefont{M.-W.} \bibnamefont{Chu}}, \bibnamefont{and}
  \bibinfo{author}{\bibfnamefont{C.~L.} \bibnamefont{Chien}},
  \bibinfo{journal}{Science} \textbf{\bibinfo{volume}{366}},
  \bibinfo{pages}{238} (\bibinfo{year}{2019}), ISSN \bibinfo{issn}{0036-8075,
  1095-9203},
  \urlprefix\url{http://www.sciencemag.org/lookup/doi/10.1126/science.aau6539}.

\bibitem[{\citenamefont{Xu et~al.}(2020)\citenamefont{Xu, Li, and
  Chien}}]{xu_spin-triplet_2020}
\bibinfo{author}{\bibfnamefont{X.}~\bibnamefont{Xu}},
  \bibinfo{author}{\bibfnamefont{Y.}~\bibnamefont{Li}}, \bibnamefont{and}
  \bibinfo{author}{\bibfnamefont{C.}~\bibnamefont{Chien}},
  \bibinfo{journal}{Physical Review Letters} \textbf{\bibinfo{volume}{124}},
  \bibinfo{pages}{167001} (\bibinfo{year}{2020}),
  \urlprefix\url{https://link.aps.org/doi/10.1103/PhysRevLett.124.167001}.

\bibitem[{\citenamefont{Hua et~al.}(2023)\citenamefont{Hua, Dumitrescu, and
  Halász}}]{Hua2023}
\bibinfo{author}{\bibfnamefont{C.}~\bibnamefont{Hua}},
  \bibinfo{author}{\bibfnamefont{E.}~\bibnamefont{Dumitrescu}},
  \bibnamefont{and} \bibinfo{author}{\bibfnamefont{G.~B.}
  \bibnamefont{Halász}}, \bibinfo{journal}{Physical Review B}
  \textbf{\bibinfo{volume}{107}}, \bibinfo{pages}{214503}
  (\bibinfo{year}{2023}), ISSN \bibinfo{issn}{2469-9969}.

\bibitem[{\citenamefont{Xu et~al.}(2024)\citenamefont{Xu, Li, and
  Chien}}]{Xu2024}
\bibinfo{author}{\bibfnamefont{X.}~\bibnamefont{Xu}},
  \bibinfo{author}{\bibfnamefont{Y.}~\bibnamefont{Li}}, \bibnamefont{and}
  \bibinfo{author}{\bibfnamefont{C.}~\bibnamefont{Chien}},
  \bibinfo{journal}{Physical Review Letters} \textbf{\bibinfo{volume}{132}},
  \bibinfo{pages}{056001} (\bibinfo{year}{2024}), ISSN
  \bibinfo{issn}{1079-7114}.

\bibitem[{\citenamefont{Li et~al.}(2024)\citenamefont{Li, Xu, Lee, and
  Chien}}]{li_unconventional_2024}
\bibinfo{author}{\bibfnamefont{Y.}~\bibnamefont{Li}},
  \bibinfo{author}{\bibfnamefont{X.}~\bibnamefont{Xu}},
  \bibinfo{author}{\bibfnamefont{S.-P.} \bibnamefont{Lee}}, \bibnamefont{and}
  \bibinfo{author}{\bibfnamefont{C.~L.} \bibnamefont{Chien}},
  \bibinfo{journal}{Physical Review B} \textbf{\bibinfo{volume}{109}},
  \bibinfo{pages}{L060504} (\bibinfo{year}{2024}), ISSN
  \bibinfo{issn}{2469-9950, 2469-9969},
  \urlprefix\url{https://link.aps.org/doi/10.1103/PhysRevB.109.L060504}.

\bibitem[{\citenamefont{Xu et~al.}(2026)\citenamefont{Xu, Qin, Shen, Huang,
  Zhou, Wang, and Li}}]{xu_signature_2025}
\bibinfo{author}{\bibfnamefont{X.}~\bibnamefont{Xu}},
  \bibinfo{author}{\bibfnamefont{W.}~\bibnamefont{Qin}},
  \bibinfo{author}{\bibfnamefont{Y.}~\bibnamefont{Shen}},
  \bibinfo{author}{\bibfnamefont{Z.}~\bibnamefont{Huang}},
  \bibinfo{author}{\bibfnamefont{Z.}~\bibnamefont{Zhou}},
  \bibinfo{author}{\bibfnamefont{Z.}~\bibnamefont{Wang}}, \bibnamefont{and}
  \bibinfo{author}{\bibfnamefont{Y.}~\bibnamefont{Li}}, \bibinfo{journal}{Phys.
  Rev. Lett.} \textbf{\bibinfo{volume}{137}}, \bibinfo{pages}{086002}
  (\bibinfo{year}{2026}),
  \urlprefix\url{https://link.aps.org/doi/10.1103/qv9w-x78v}.

\bibitem[{\citenamefont{Almoalem et~al.}(2024)\citenamefont{Almoalem, Feldman,
  Mangel, Shlafman, Yaish, Fischer, Moshe, Ruhman, and
  Kanigel}}]{almoalem_observation_2024}
\bibinfo{author}{\bibfnamefont{A.}~\bibnamefont{Almoalem}},
  \bibinfo{author}{\bibfnamefont{I.}~\bibnamefont{Feldman}},
  \bibinfo{author}{\bibfnamefont{I.}~\bibnamefont{Mangel}},
  \bibinfo{author}{\bibfnamefont{M.}~\bibnamefont{Shlafman}},
  \bibinfo{author}{\bibfnamefont{Y.~E.} \bibnamefont{Yaish}},
  \bibinfo{author}{\bibfnamefont{M.~H.} \bibnamefont{Fischer}},
  \bibinfo{author}{\bibfnamefont{M.}~\bibnamefont{Moshe}},
  \bibinfo{author}{\bibfnamefont{J.}~\bibnamefont{Ruhman}}, \bibnamefont{and}
  \bibinfo{author}{\bibfnamefont{A.}~\bibnamefont{Kanigel}},
  \bibinfo{journal}{Nature Communications} \textbf{\bibinfo{volume}{15}},
  \bibinfo{pages}{4623} (\bibinfo{year}{2024}), ISSN \bibinfo{issn}{2041-1723},
  \urlprefix\url{https://www.nature.com/articles/s41467-024-48260-x}.

\bibitem[{\citenamefont{Wan et~al.}(2024)\citenamefont{Wan, Qiu, Ren, Qian, Li,
  Xu, Zhou, Zhou, Zhou, Wang et~al.}}]{wan_unconventional_2024}
\bibinfo{author}{\bibfnamefont{Z.}~\bibnamefont{Wan}},
  \bibinfo{author}{\bibfnamefont{G.}~\bibnamefont{Qiu}},
  \bibinfo{author}{\bibfnamefont{H.}~\bibnamefont{Ren}},
  \bibinfo{author}{\bibfnamefont{Q.}~\bibnamefont{Qian}},
  \bibinfo{author}{\bibfnamefont{Y.}~\bibnamefont{Li}},
  \bibinfo{author}{\bibfnamefont{D.}~\bibnamefont{Xu}},
  \bibinfo{author}{\bibfnamefont{J.}~\bibnamefont{Zhou}},
  \bibinfo{author}{\bibfnamefont{J.}~\bibnamefont{Zhou}},
  \bibinfo{author}{\bibfnamefont{B.}~\bibnamefont{Zhou}},
  \bibinfo{author}{\bibfnamefont{L.}~\bibnamefont{Wang}}, \bibnamefont{et~al.},
  \bibinfo{journal}{Nature} \textbf{\bibinfo{volume}{632}}, \bibinfo{pages}{69}
  (\bibinfo{year}{2024}), ISSN \bibinfo{issn}{1476-4687},
  \urlprefix\url{https://www.nature.com/articles/s41586-024-07625-4}.

\bibitem[{\citenamefont{Tokuda et~al.}(2025)\citenamefont{Tokuda, Matsumoto,
  Maeda, Higashihara, Nakao, Watanabe, Lee, Nakamura, Maeda, Jiang
  et~al.}}]{Tokuda2025}
\bibinfo{author}{\bibfnamefont{M.}~\bibnamefont{Tokuda}},
  \bibinfo{author}{\bibfnamefont{F.}~\bibnamefont{Matsumoto}},
  \bibinfo{author}{\bibfnamefont{N.}~\bibnamefont{Maeda}},
  \bibinfo{author}{\bibfnamefont{T.}~\bibnamefont{Higashihara}},
  \bibinfo{author}{\bibfnamefont{M.}~\bibnamefont{Nakao}},
  \bibinfo{author}{\bibfnamefont{M.}~\bibnamefont{Watanabe}},
  \bibinfo{author}{\bibfnamefont{S.}~\bibnamefont{Lee}},
  \bibinfo{author}{\bibfnamefont{R.}~\bibnamefont{Nakamura}},
  \bibinfo{author}{\bibfnamefont{M.}~\bibnamefont{Maeda}},
  \bibinfo{author}{\bibfnamefont{N.}~\bibnamefont{Jiang}},
  \bibnamefont{et~al.}, \bibinfo{journal}{Science Advances}
  \textbf{\bibinfo{volume}{11}} (\bibinfo{year}{2025}), ISSN
  \bibinfo{issn}{2375-2548}.

\bibitem[{\citenamefont{Wang et~al.}(2025)\citenamefont{Wang, Maccari, Feng,
  Wu, Peng, Law, Zhao, Szabo, Schnyder, Kang et~al.}}]{wang_switchable_2025}
\bibinfo{author}{\bibfnamefont{S.}~\bibnamefont{Wang}},
  \bibinfo{author}{\bibfnamefont{I.}~\bibnamefont{Maccari}},
  \bibinfo{author}{\bibfnamefont{X.}~\bibnamefont{Feng}},
  \bibinfo{author}{\bibfnamefont{Z.-N.} \bibnamefont{Wu}},
  \bibinfo{author}{\bibfnamefont{J.-P.} \bibnamefont{Peng}},
  \bibinfo{author}{\bibfnamefont{K.~T.} \bibnamefont{Law}},
  \bibinfo{author}{\bibfnamefont{Y.~X.} \bibnamefont{Zhao}},
  \bibinfo{author}{\bibfnamefont{A.}~\bibnamefont{Szabo}},
  \bibinfo{author}{\bibfnamefont{A.}~\bibnamefont{Schnyder}},
  \bibinfo{author}{\bibfnamefont{N.}~\bibnamefont{Kang}}, \bibnamefont{et~al.},
  \emph{\bibinfo{title}{Switchable half-quantum flux states in a ring of the
  kagome superconductor {CsV}\$\_3\${Sb}\$\_5\$}} (\bibinfo{year}{2025}),
  \bibinfo{note}{arXiv:2512.10010 [cond-mat]},
  \urlprefix\url{http://arxiv.org/abs/2512.10010}.

\bibitem[{\citenamefont{Tinkham}(1963)}]{Tinkham1963}
\bibinfo{author}{\bibfnamefont{M.}~\bibnamefont{Tinkham}},
  \bibinfo{journal}{Physical Review} \textbf{\bibinfo{volume}{129}},
  \bibinfo{pages}{2413} (\bibinfo{year}{1963}), ISSN \bibinfo{issn}{0031-899X}.

\bibitem[{\citenamefont{TINKHAM}(1964)}]{TINKHAM1964}
\bibinfo{author}{\bibfnamefont{M.}~\bibnamefont{TINKHAM}},
  \bibinfo{journal}{Reviews of Modern Physics} \textbf{\bibinfo{volume}{36}},
  \bibinfo{pages}{268} (\bibinfo{year}{1964}), ISSN \bibinfo{issn}{0034-6861}.

\bibitem[{\citenamefont{Groff and Parks}(1968)}]{groff_fluxoid_1968}
\bibinfo{author}{\bibfnamefont{R.~P.} \bibnamefont{Groff}} \bibnamefont{and}
  \bibinfo{author}{\bibfnamefont{R.~D.} \bibnamefont{Parks}},
  \bibinfo{journal}{Physical Review} \textbf{\bibinfo{volume}{176}},
  \bibinfo{pages}{567} (\bibinfo{year}{1968}), ISSN \bibinfo{issn}{0031-899X},
  \urlprefix\url{https://link.aps.org/doi/10.1103/PhysRev.176.567}.

\bibitem[{\citenamefont{Sochnikov
  et~al.}(2010{\natexlab{b}})\citenamefont{Sochnikov, Shaulov, Yeshurun,
  Logvenov, and Božović}}]{sochnikov_oscillatory_2010}
\bibinfo{author}{\bibfnamefont{I.}~\bibnamefont{Sochnikov}},
  \bibinfo{author}{\bibfnamefont{A.}~\bibnamefont{Shaulov}},
  \bibinfo{author}{\bibfnamefont{Y.}~\bibnamefont{Yeshurun}},
  \bibinfo{author}{\bibfnamefont{G.}~\bibnamefont{Logvenov}}, \bibnamefont{and}
  \bibinfo{author}{\bibfnamefont{I.}~\bibnamefont{Božović}},
  \bibinfo{journal}{Physical Review B} \textbf{\bibinfo{volume}{82}},
  \bibinfo{pages}{094513} (\bibinfo{year}{2010}{\natexlab{b}}), ISSN
  \bibinfo{issn}{1098-0121, 1550-235X},
  \urlprefix\url{https://link.aps.org/doi/10.1103/PhysRevB.82.094513}.

\bibitem[{\citenamefont{Langer and Ambegaokar}(1967)}]{langer_Ambegaokar_1967}
\bibinfo{author}{\bibfnamefont{J.}~\bibnamefont{Langer}} \bibnamefont{and}
  \bibinfo{author}{\bibfnamefont{V.}~\bibnamefont{Ambegaokar}},
  \bibinfo{journal}{Physical Review Letters} \textbf{\bibinfo{volume}{164}},
  \bibinfo{pages}{498} (\bibinfo{year}{1967}),
  \urlprefix\url{https://doi.org/10.1103/PhysRev.164.498}.

\bibitem[{\citenamefont{McCumber}(1968)}]{mccumber_intrinsic_1968}
\bibinfo{author}{\bibfnamefont{D.~E.} \bibnamefont{McCumber}},
  \bibinfo{journal}{Physical Review} \textbf{\bibinfo{volume}{172}},
  \bibinfo{pages}{427} (\bibinfo{year}{1968}),
  \urlprefix\url{https://doi.org/10.1103/PhysRev.172.427}.

\bibitem[{\citenamefont{Moshchalkov et~al.}(1995)\citenamefont{Moshchalkov,
  Gielen, Strunk, Jonckheere, Qiu, Haesendonck, and
  Bruynseraede}}]{moshchalkov_effect_1995}
\bibinfo{author}{\bibfnamefont{V.~V.} \bibnamefont{Moshchalkov}},
  \bibinfo{author}{\bibfnamefont{L.}~\bibnamefont{Gielen}},
  \bibinfo{author}{\bibfnamefont{C.}~\bibnamefont{Strunk}},
  \bibinfo{author}{\bibfnamefont{R.}~\bibnamefont{Jonckheere}},
  \bibinfo{author}{\bibfnamefont{X.}~\bibnamefont{Qiu}},
  \bibinfo{author}{\bibfnamefont{C.~V.} \bibnamefont{Haesendonck}},
  \bibnamefont{and}
  \bibinfo{author}{\bibfnamefont{Y.}~\bibnamefont{Bruynseraede}},
  \bibinfo{journal}{Nature} \textbf{\bibinfo{volume}{373}},
  \bibinfo{pages}{319} (\bibinfo{year}{1995}), ISSN \bibinfo{issn}{1476-4687},
  \urlprefix\url{https://www.nature.com/articles/373319a0}.

\bibitem[{\citenamefont{see SUPPLEMENTAL MATERIAL~for details.}(2026)}]{supple}
\bibinfo{author}{\bibnamefont{see SUPPLEMENTAL MATERIAL~for details.}}
  (\bibinfo{year}{2026}).

\bibitem[{\citenamefont{Powell et~al.}(2025)\citenamefont{Powell, Kuang,
  Hawkins-Pottier, Jalil, Birkbeck, Jiang, Kim, Zou, Komrakova, Haigh
  et~al.}}]{Powell2025}
\bibinfo{author}{\bibfnamefont{L.}~\bibnamefont{Powell}},
  \bibinfo{author}{\bibfnamefont{W.}~\bibnamefont{Kuang}},
  \bibinfo{author}{\bibfnamefont{G.}~\bibnamefont{Hawkins-Pottier}},
  \bibinfo{author}{\bibfnamefont{R.}~\bibnamefont{Jalil}},
  \bibinfo{author}{\bibfnamefont{J.}~\bibnamefont{Birkbeck}},
  \bibinfo{author}{\bibfnamefont{Z.}~\bibnamefont{Jiang}},
  \bibinfo{author}{\bibfnamefont{M.}~\bibnamefont{Kim}},
  \bibinfo{author}{\bibfnamefont{Y.}~\bibnamefont{Zou}},
  \bibinfo{author}{\bibfnamefont{S.}~\bibnamefont{Komrakova}},
  \bibinfo{author}{\bibfnamefont{S.}~\bibnamefont{Haigh}},
  \bibnamefont{et~al.}, \bibinfo{journal}{Nature Communications}
  \textbf{\bibinfo{volume}{16}}, \bibinfo{pages}{291} (\bibinfo{year}{2025}),
  ISSN \bibinfo{issn}{2041-1723},
  \urlprefix\url{https://www.nature.com/articles/s41467-024-54867-x}.

\bibitem[{\citenamefont{Parks and Little}(1964)}]{parks_fluxoid_1964}
\bibinfo{author}{\bibfnamefont{R.~D.} \bibnamefont{Parks}} \bibnamefont{and}
  \bibinfo{author}{\bibfnamefont{W.~A.} \bibnamefont{Little}},
  \bibinfo{journal}{Physical Review} \textbf{\bibinfo{volume}{133}},
  \bibinfo{pages}{A97} (\bibinfo{year}{1964}), ISSN \bibinfo{issn}{0031-899X},
  \urlprefix\url{https://link.aps.org/doi/10.1103/PhysRev.133.A97}.

\bibitem[{\citenamefont{Lukens et~al.}(1970)\citenamefont{Lukens, Warburton,
  and Webb}}]{lukens_onset_1970}
\bibinfo{author}{\bibfnamefont{J.~E.} \bibnamefont{Lukens}},
  \bibinfo{author}{\bibfnamefont{R.~J.} \bibnamefont{Warburton}},
  \bibnamefont{and} \bibinfo{author}{\bibfnamefont{W.~W.} \bibnamefont{Webb}},
  \bibinfo{journal}{Physical Review Letters} \textbf{\bibinfo{volume}{25}},
  \bibinfo{pages}{1180} (\bibinfo{year}{1970}),
  \urlprefix\url{https://doi.org/10.1103/PhysRevLett.25.1180}.

\bibitem[{\citenamefont{Giordano}(1988)}]{giordano_evidence_1988}
\bibinfo{author}{\bibfnamefont{N.}~\bibnamefont{Giordano}},
  \bibinfo{journal}{Physical Review Letters} \textbf{\bibinfo{volume}{61}},
  \bibinfo{pages}{2137} (\bibinfo{year}{1988}),
  \urlprefix\url{https://doi.org/10.1103/PhysRevLett.61.2137}.

\bibitem[{\citenamefont{Arutyunov et~al.}(2008)\citenamefont{Arutyunov,
  Golubev, and Zaikin}}]{arutyunov_superconductivity_2008}
\bibinfo{author}{\bibfnamefont{K.~Y.} \bibnamefont{Arutyunov}},
  \bibinfo{author}{\bibfnamefont{D.~S.} \bibnamefont{Golubev}},
  \bibnamefont{and} \bibinfo{author}{\bibfnamefont{A.~D.}
  \bibnamefont{Zaikin}}, \bibinfo{journal}{Physics Reports}
  \textbf{\bibinfo{volume}{464}}, \bibinfo{pages}{1} (\bibinfo{year}{2008}),
  \urlprefix\url{https://doi.org/10.1016/j.physrep.2008.04.009}.

\end{thebibliography}

\clearpage

\begin{figure}
\centering
\includegraphics[width=17cm]{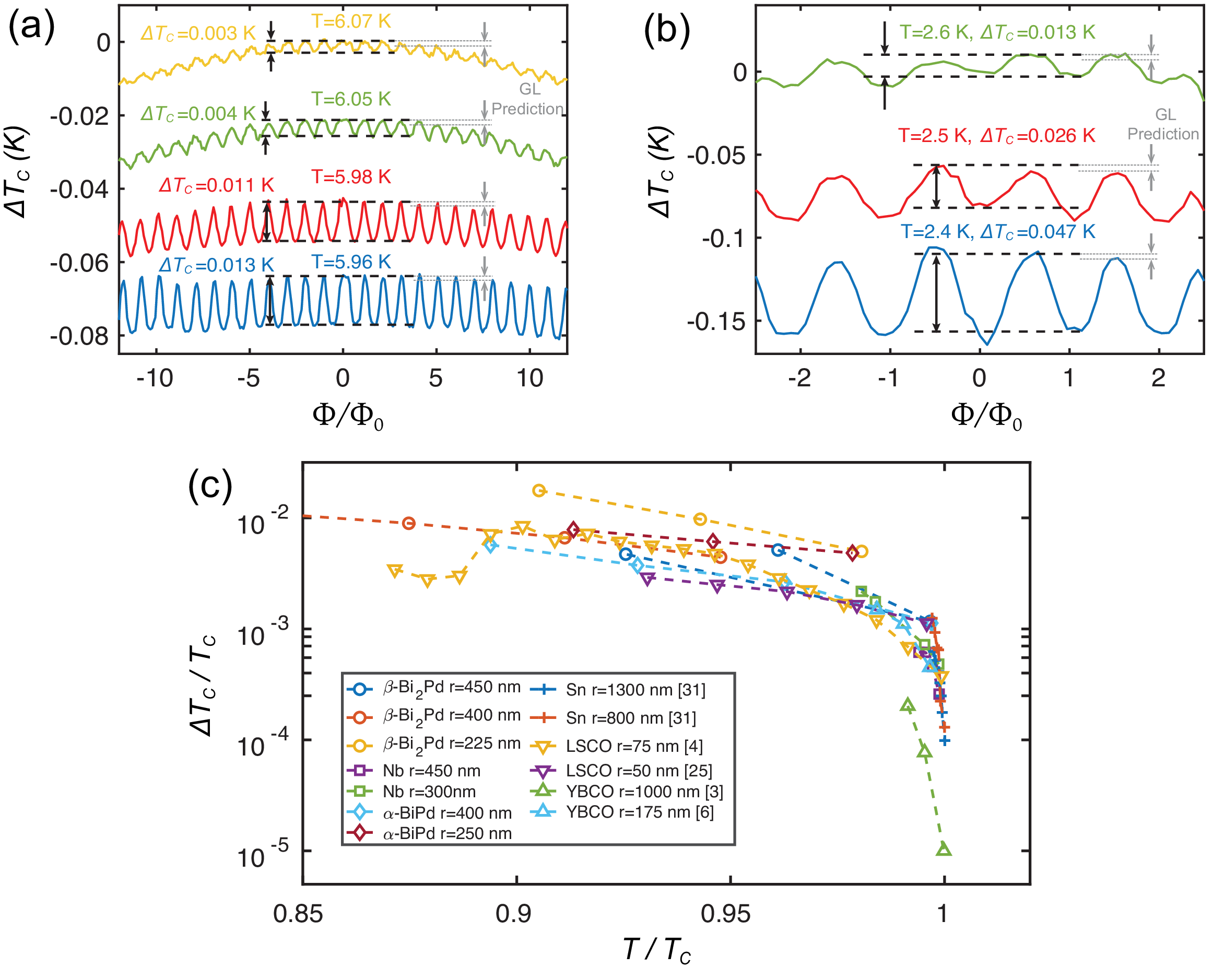}
\caption{Little--Parks oscillation curves measured at different temperatures for (a) the Nb ring ($r=300$~nm) and (b) the $\beta$-Bi$_{2}$Pd ring ($r=225$~nm); the dashed black lines indicate the measured amplitude and the dashed gray lines indicate the amplitude given by the GL theory. (c) Normalized Little--Parks oscillation amplitudes, $\frac{\Delta T_C}{T_C}$, plotted as a function of $\frac{T}{T_C}$ for various superconducting rings, including those reported in previous studies.}
\end{figure}

\begin{figure}
\centering
\includegraphics[width=16cm]{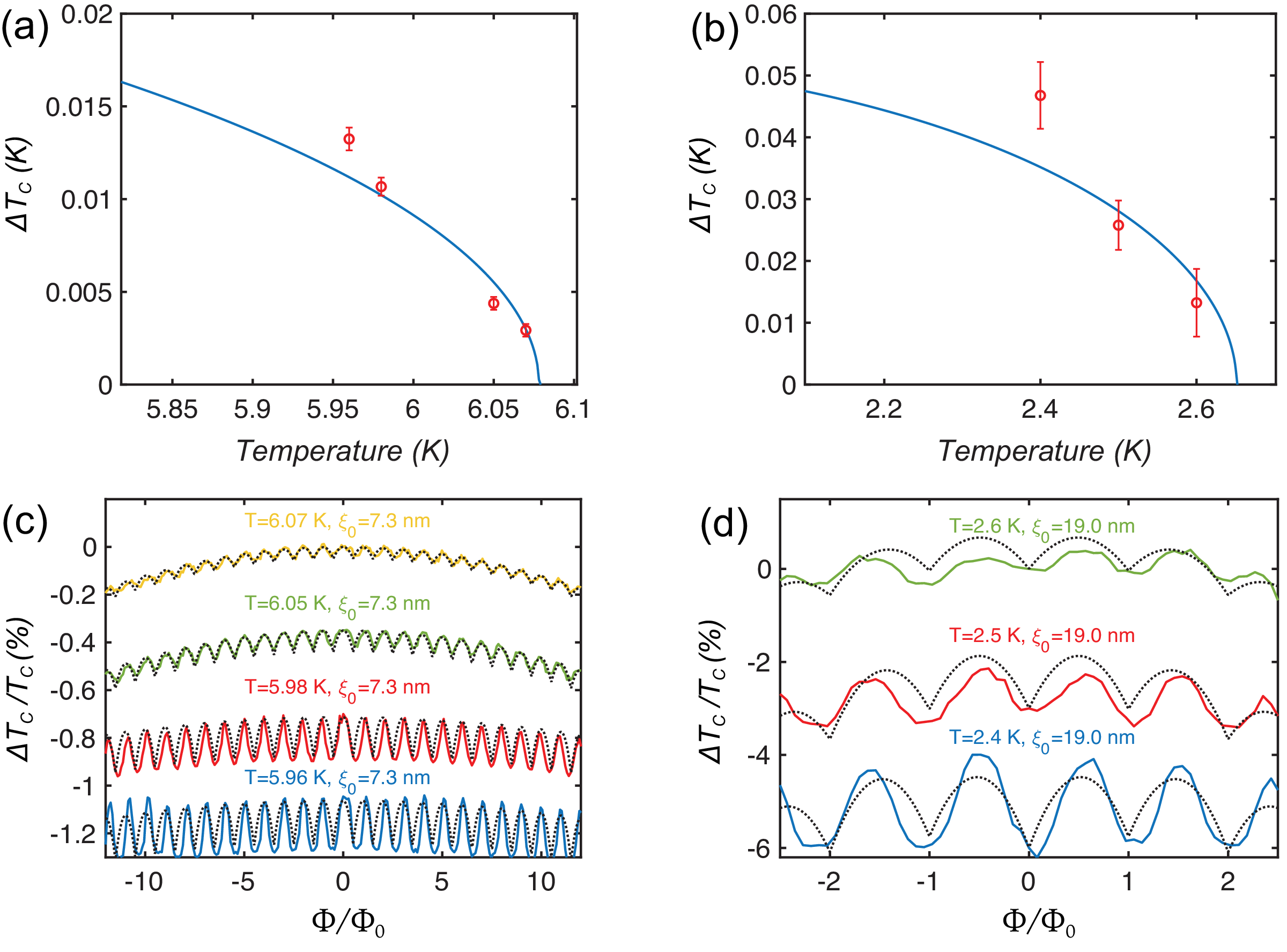}
\caption{Little--Parks oscillation amplitudes, $\frac{\Delta T_C}{T_C}$ of (a) the Nb ring ($r=300$~nm) and (b) the $\beta$-Bi$_{2}$Pd ring ($r=225$~nm) measured at different temperatures. The solid blue lines are fits to $\frac{\Delta T_C}{T_C}=\chi (t)\frac{\xi_0}{4r}$, yielding zero-temperature coherence lengths of 7.3~nm for Nb and 19.0~nm for $\beta$-Bi$_{2}$Pd. Panels (c) and (d) show the corresponding Little--Parks oscillation curves of the Nb ring and $\beta$-Bi$_{2}$Pd rings, respectively, fitted using Eq.~(4) with the extracted $\xi_0$. }
\end{figure}

\begin{figure}
\centering
\includegraphics[width=14cm]{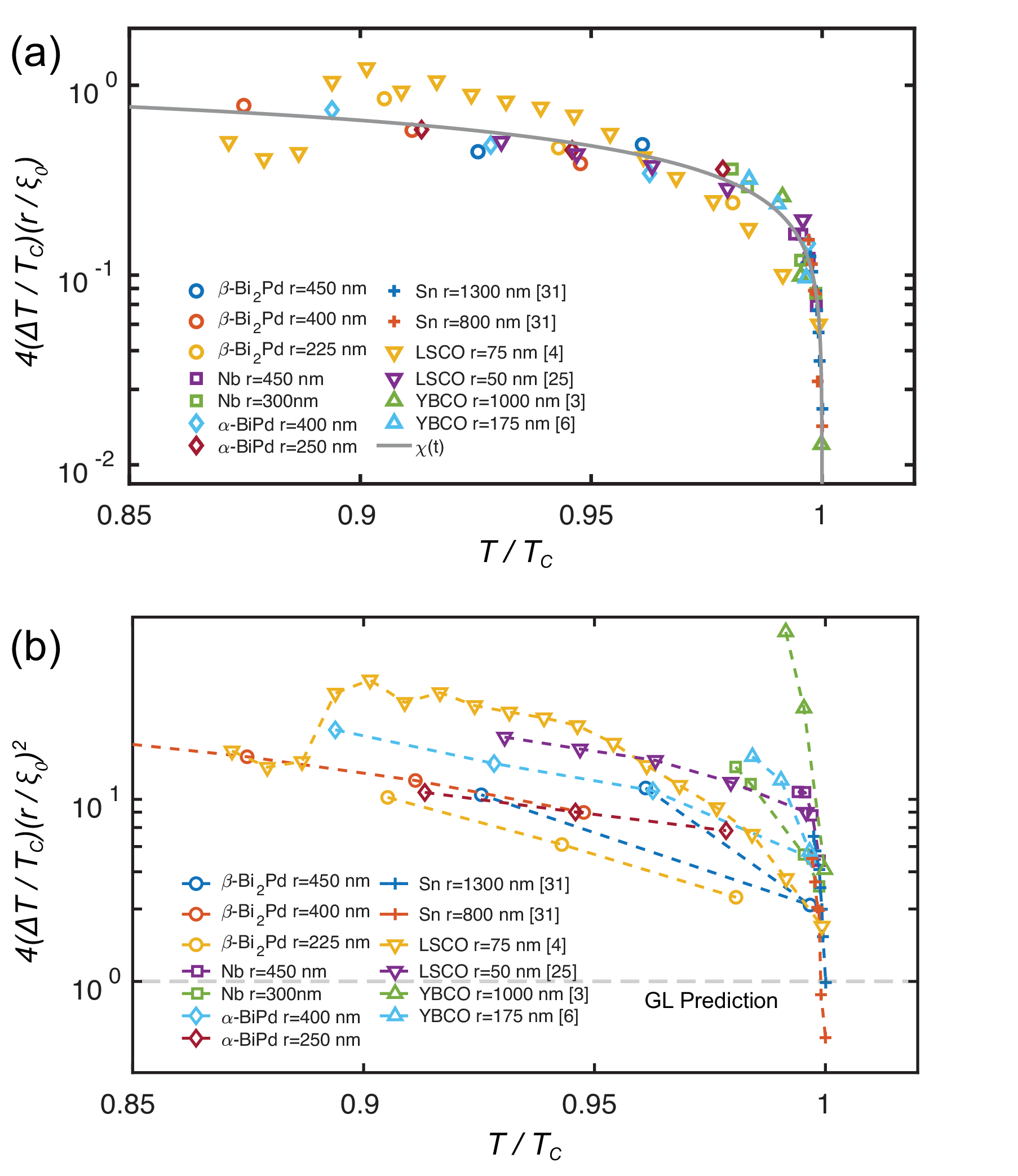}
\caption{(a) Normalized Little--Parks oscillation amplitudes, $\frac{\Delta T_C}{T_C}\frac{4r}{\xi_0}$, plotted as a function of $T/T_C$ for various superconducting rings. The solid gray line shows $\chi(t)$ as a function of $T/T_C$. 
(b) The same data normalized according to the conventional GL scaling, $\frac{\Delta T_C}{T_C}\frac{4r^2}{\xi_0^2}$.
The dashed gray line marks the GL prediction of unity.
}
\end{figure}
\clearpage

\textbf{ACKNOWLEDGMENTS}

This work was supported by National Natural Science Foundation of China (Grants No. 12474153, No. 12474134), Innovation Program for Quantum Science
and Technology: Grant No. 2021ZD0302800, Guangdong Provincial Quantum Science Strategic Initiative (No. GDZX2301007, No. GDZX2401003, No. GDZX2501005), Research Grants Council, University Grants Committee of the Hong Kong SAR (Grant No. CUHK 24302822, No. CUHK 14306823), CUHK Direct Grants (No. 4053668, No. 4053727). X. X acknowledges the support by the startup funding of the Quantum Science Center under Grant No. QD2401002.

\textbf{Competing interests:}  
The authors declare no competing interests.

\textbf{Data availability statement:}
The datasets generated during and/or analyzed during the current study are available from the corresponding author on reasonable request.
A copy shall be made available on a third party repository at publication.

\clearpage


\begingroup
\renewcommand{\theequation}{S\arabic{equation}}
\renewcommand{\thefigure}{S\arabic{figure}}
\renewcommand{\thetable}{S\arabic{table}}
\setcounter{equation}{0}
\setcounter{figure}{0}
\setcounter{table}{0}
\begin{center}
{\large\bfseries Supplemental Material}
\end{center}
\vspace{1em}

\makeatletter
\setcounter{secnumdepth}{3}
\renewcommand{\thesection}{\Roman{section}}
\renewcommand{\@seccntformat}[1]{\csname the#1\endcsname.\quad}
\makeatother

\section{Conversion from resistance oscillation to $\Delta T_C$ oscillation}

In the Little-Parks experiments, the resistance is measured as a function of magnetic flux at a fixed temperature close to $T_C$.
For a measurement performed at a temperature $T_0$, the resistance $R(T_0, \Phi)$ is mapped onto the zero-flux $R(T_0, 0)$ transition curve to find the $\Delta T_C$. More specifically,
\begin{equation}
    \Delta T_C(\Phi)=T_0-T^{(0)}[R(T_0,\Phi)].
\end{equation}
where $T^{(0)}(R)$ denotes the inverse function of the zero-flux $R-T$ curve. This conversion relies only on the $R-T$ curve to be rigid locally around $T_0$, a valid assumption given that $\Delta R$ is small. 
Details of this conversion procedure can also be found in Ref.~17.

Conventionally, the Little-Parks oscillation is derived from the linearized Ginzburg-Landau equation that predicts a globally rigid shift of the $R-T$ curve shown as the red curve in Fig. S1. However, experiments have shown temperature-dependent $\Delta T_C$, indicating that while the local rigidity curve is a valid assumption, $R-T$ does not globally shift with rigidity.
After employing the thermally activated phase slip (TAPS), the  $R-T$ curve would shift as the yellow curve shown in Fig S1, giving rise to a temperature dependent $\Delta T_C$. 

\begin{figure}[htp]
\centering
\includegraphics[width=8cm]{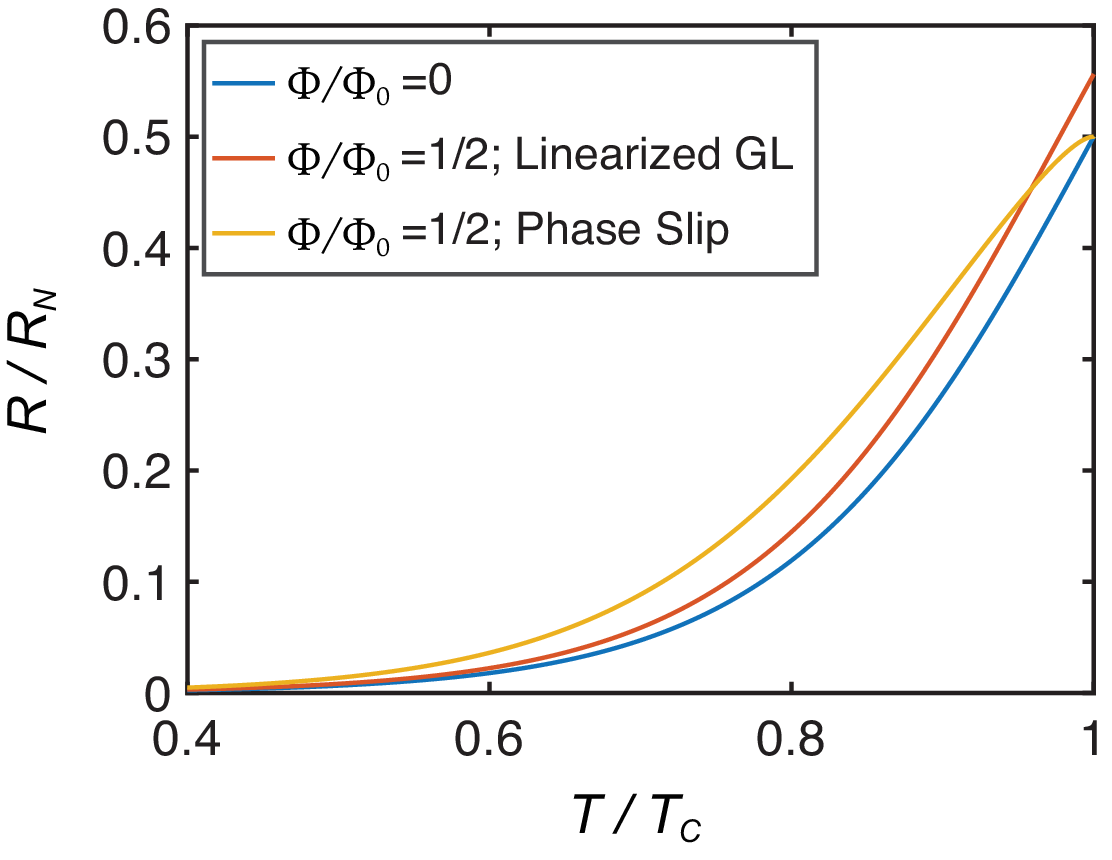}
\caption{The blue curve is $R-T$ without external magnetic field. The red curve is $R-T$ at $\Phi=1/2\Phi_0$ after being rigidly shifted, as given by linearized GL following Eq. 1. The yellow curve is $R-T$ at $\Phi=1/2\Phi_0$ given by TAPS following Eq.(4) in the main text. 
}
\end{figure}

\section{Zero-temperature coherence lengths estimated from $H_{C2}$}

The zero-temperature Ginzburg-Landau coherence length ($\xi_0$) can be estimated from the temperature dependence of the upper critical field $H_{C2}$.
Near $T_C$, the upper critical field is given by

\begin{equation}
    \mu_0 H_{C2}(T)=\frac{\Phi_0}{2\pi\xi^2(T)}=\frac{\Phi_0}{2\pi\xi_0^2}\left(1-\frac{T}{T_C}\right).
\end{equation}

Taking the derivative with respect to temperature gives
$d(\mu_0 H_{C2})/dT$=-$\Phi_0/(2\pi \xi_0^2 T_C)$.
Therefore, $\xi_0$ can be obtained from the slope of $\mu_0 H_{C2}$\ near $T_C$ as

\begin{equation}
    \xi_0=\sqrt{-\frac{\Phi_0}{2\pi T_C \left[\frac{d(\mu_0H_{C2})}{dT}\right]}}.
\end{equation}

The $d(\mu_0H_{C2})/dT$ term is determined from a linear fit to the upper-critical field data close to $T_C$.
The measured upper-critical fields of the 40~nm-thick Nb thin film and the 50~nm-thick $\beta$-Bi$_{2}$Pd film are shown in Fig.~S2.
The Nb and $\beta$-Bi$_{2}$Pd devices discussed in the main text were fabricated from the films with the same respective thicknesses.
The corresponding zero-temperature Ginzburg-Landau coherence lengths were extracted from the linear fits using Eq.~S3.
The obtained $\xi_0$ are 7.0~nm for Nb and 14.3~nm for $\beta$-Bi$_{2}$Pd, which are close to values reported in the literature.

\begin{figure}
\centering
\includegraphics[width=16cm]{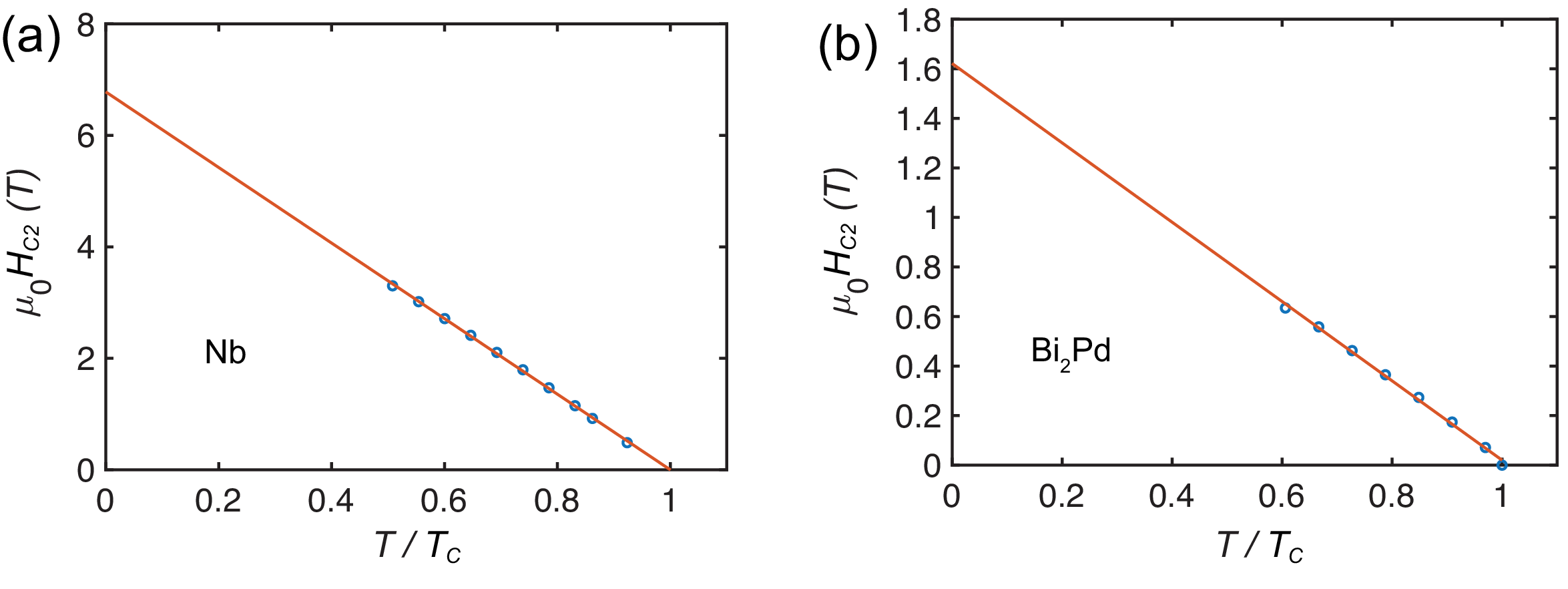}
\caption{ $H_{C2}$ of the $40$ nm-thick Nb film (a) and the $50$ nm-thick $\beta$-Bi$_{2}$Pd film (b) measured at various temperatures near $T_C$. The red line is the linear fit to the experimental data. }
\end{figure}

\section{Derivation of the new thermal-fluctuation-corrected equation}

In a thin superconducting wire, thermal fluctuations can momentarily and locally suppress superconductivity, then transition to a state that changes phase by $2\pi$, leading to a small change in voltage given by the Josephson relationship
\begin{equation}
    (2e/\hbar)\Delta V = \partial(\Delta\phi)/\partial t .
\end{equation}
Thus, the resistance $R$ measured across the wire is proportional to the TAPS rate $\Gamma$, which follows the relationship
\begin{equation}
    R \propto \Gamma = \exp\left[-\Delta F(T,\Phi)/k_BT\right],
\end{equation}
where $\Delta F$ is the free energy barrier for TAPS to occur. 

There are three primary contributions to $\Delta F$. The first contribution $\Delta F_0$ is the energy required to suppress the order parameter $\psi$ to zero. $\Delta F_0$ is calculated in the LAMH theory and expressed as the equation
\begin{equation}
\Delta F_0(T)
=
\frac{8\sqrt{2}}{3}
\frac{H_c^2}{8\pi}A\xi,
\label{eq:zfeb}
\end{equation}
with $\xi=\xi_0(1-T/T_c)^{1/2}$, $\alpha(T)=(\hbar^2/(2m^*\xi_0^2T_c))(T-T_c)=\alpha_0(T-T_c)$, $H_c^2=4\pi\alpha^2(T)/\beta$ and $A=w\,d$ being the cross-sectional area of the wire. $\Delta F_0$ is the energy the system consumes to locally suppress the order parameter, so the larger the $\Delta F_0$ value, the lower the phase-slip rate and the lower the resistance, and vice versa; hence
\begin{equation}
R\propto \exp\left[-\Delta F_0(T)/k_BT\right].
\end{equation}

The second contribution arises from the kinetic energy of the circulating supercurrent induced by the magnetic flux threading the superconducting ring. The magnetic flux generates a finite gauge-invariant phase gradient, giving rise to an additional kinetic-energy contribution to the GL free energy. Within the framework of TAPS, the resulting circulating supercurrent modifies the free-energy landscape and generally lowers the activation barrier for phase slips that relax the supercurrent, thereby enhancing the phase-slip rate.
Qualitatively, the larger the kinetic energy, the higher the phase-slip rate and the higher the resistance, and vise versa. Incorporating this contribution and assuming that the kinetic-energy correction remains small, as is typically the case in Little–Parks experiments, the resistance can be approximated as
\begin{equation}
R \approx \Omega\,\exp\left[\frac{-\Delta F_0(T)+\Delta F_{\mathrm{kin}}(T,\Phi)}{k_BT}\right].
\end{equation}
Here $\Omega$ is a device-specific constant and $\Delta F_{\mathrm{kin}}(T,\Phi)$ denotes the kinetic energy correction given by 
\begin{equation}
\Delta F_{\mathrm{kin}}=\frac{\hbar^2}{2m^*}\int dx^3 \left(\textbf{D}\psi\right)^2,
\end{equation}
with $\psi$ representing the superconducting order parameter.
For a ring with radius $r$ and arm-width $w$, a uniform magnetic field and $\psi=\sqrt{-\alpha(T)/\beta}\,e^{in\theta}$, the kinetic energy is derived as
\begin{equation}\label{Fkin}
\Delta F_{\mathrm{kin}}=2\pi\frac{\alpha_0^2\,\xi_0^2 T_C^2}{\beta} \frac{w\,d}{r}\,\left[\left(\frac{w}{r}\right)^2\left(\frac{\Phi}{\Phi_0}\right)^2+\left(
n-\frac{\Phi}{\Phi_0}
\right)^2\right](1-t)\,,
\end{equation}
with the reduced temperature $t=T/T_C$. 

The third contribution arises from the externally applied bias current used to measure the resistance, which flows through the two arms of the superconducting ring. The bias current in Little-Parks measurements typically ranges from 1 $\mu$A to 10 $\mu$A. For a ring geometry with $r=500$~nm and $w=200$~nm, the inductance is roughly $L\approx 1\times10^{-10}$~H, yielding a circulating current of about $I\approx 1$~mA. Since the applied bias current is much smaller than the circulating supercurrent, we neglect its contribution to the modification of the kinetic energy in the present analysis. This approximation is further supported by our experimental measurements performed at different bias currents, which exhibit nearly identical oscillation amplitudes of $T_c$, indicating that the influence of the bias current is negligible under the experimental conditions.

Based on the expression of resistance, we can derive the oscillation magnitude of $T_c$.
In experiments, $\Delta T_c=T_c(H)-T_c(0)$ is determined by identifying the resistance with and without the magnetic field, hence $R(T+\Delta T_c,\Phi)=R(T,0).$, indicating
\begin{equation}
\Delta F_0( T+\Delta T_C)-\Delta F_{\mathrm{kin}}(T+\Delta T_C,\Phi)  =  \Delta F_0(T) (1+\Delta T_C /T).
\end{equation}
Since $\Delta T_c$ is typically smaller than $T$, the free energies can be expanded perturbatively, yielding
\begin{equation}
\Delta F_0( T) -  \Delta F_{\mathrm{kin}}(T,\Phi) +\left(\frac{\partial \Delta F_0(T)}{\partial T}  -\frac{\partial \Delta F_{\mathrm{kin}}(T,\Phi)}{\partial T} \right) \Delta T_C =  \Delta F_0(T) (1+\Delta T_C /T).
\end{equation}
Rearranging this equation yields 
 \begin{equation}
 \Delta T_C  = \frac{\Delta F_{\mathrm{kin}}(T,\Phi) }{\partial (\Delta F_0(T))/\partial T- \Delta F_0( T)/T  - \partial \Delta F_{\mathrm{kin}}(T,\Phi)/\partial T} \,\,.
 \end{equation}
 The first two terms in the denominator can be directly calculated from Eq.~(\ref{eq:zfeb}), yielding
\begin{equation}\label{denominator}
    \left[\frac{\partial \Delta F_0(T)}{\partial T}
-\frac{\Delta F_0(T)}{T}
\right]=
-\frac{4\sqrt{2}}{3}
\frac{w\,d\,\xi_0\alpha_0^2 T_C}{\beta}
(1-t)^{1/2}
\left[
\frac{1-t}{t}
+
\frac{3}{2}
\right]\,.
\end{equation}
The third term in the denominator, $\partial \Delta F_{\mathrm{kin}}(T,\Phi)/\partial T$, is negligible compared with the other terms, as $\Delta F_{\mathrm{kin}}$ is suppressed by the small dimensionless factor $\xi_0/R \ll 1$ for typical Little–Parks devices. Therefore, we neglect this term in the above equation.
Combining Eqs.~(\ref{Fkin}) and (\ref{denominator}), we obtain the expression of $\Delta T_c$ in Eq.(4) in the main text as
\begin{equation}
\frac{\Delta T_C}{T_C}
=
\chi (t)\frac{\xi_0}{r}
\left[
-\left(\frac{w}{2r}\right)^2
\left(\frac{\Phi}{\Phi_0}\right)^2
-
\left(n-\frac{\Phi}{\Phi_0}\right)^2
\right] ,
\end{equation}
where $\chi(t)=(3\pi/\sqrt2)\cdot (t\sqrt{1-t}/(t+2))$.
The numerator is precisely the kinetic-energy correction $\Delta F_{\mathrm{kin}}$, which also appears on the right-hand side of the conventional Little–Parks equation derived from the linearized GL theory. Consequently, the flux dependence of both the oscillatory and background contributions retains the same functional form as in the conventional Little–Parks equation. More generally, any characteristic algebraic structures in the free energy arising from unconventional superconducting order parameters are preserved in the present formulation, thereby carrying over the corresponding features of the conventional Little–Parks equation to the newly derived expression.

\section{Little-Parks experiments on $\alpha$-BiPd ring}

In this section, we present the Little-Parks measurements on an $\alpha$-BiPd ring.
Fig.~S3(a) shows the Little-Parks oscillations measured in a ring with $r=400$~nm at different temperatures.
The oscillation amplitude exhibits a clear temperature dependence, which can be fitted by $\Delta T_C=\chi (t)\frac{\xi_0}{4r}T_C$.
Using Eq.~4, the temperature-dependent Little-Parks oscillations can be fitted with a single zero-temperature coherence length, $\xi_0=12.4$~nm. 

\begin{figure}[htp]
\centering
\includegraphics[width=16cm]{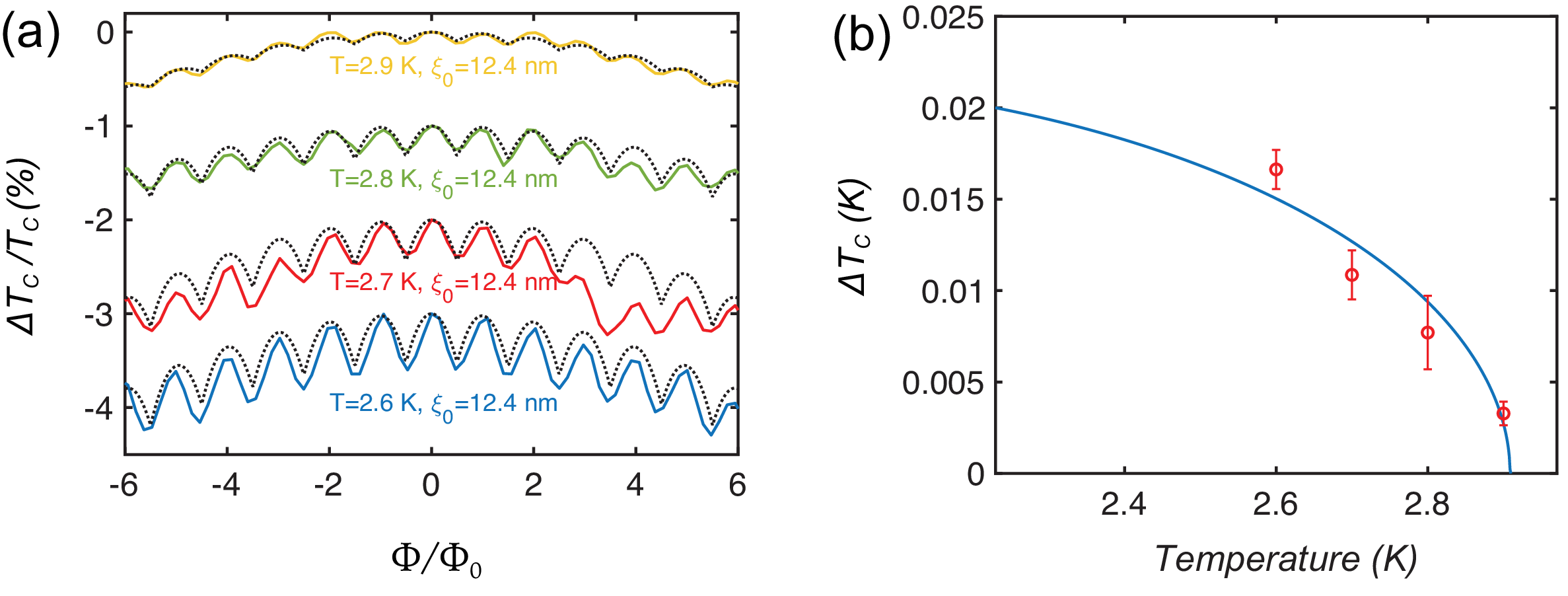}
\caption{(a) The Little-Parks oscillation curves of the $\alpha$-BiPd rings with $r=400$~nm at different temperatures. The solid lines are the experimental data and the black dashed lines are fitting lines using Eq.~4. (b) Temperature-dependent oscillation amplitudes at different temperatures. The blue line is the fitting line by $\Delta T_C=\chi (t)\frac{\xi_0}{4r}T_C$ with $\xi_0=12.5$~nm. }
\end{figure}

\section{Temperature-dependence of the Little-Parks amplitude in rings with different radii}

In this section, we show that the Little-Parks amplitude is temperature-dependent in ring devices with different radii for Nb, $\beta$-Bi$_2$Pd and $\alpha$-BiPd, demonstrating that this behavior is consistently observed in both conventional and unconventional superconducting rings. 
Fig.~S4 shows the Little-Parks oscillations measured in a Nb ring with radius $r=450$~nm at different temperatures.
The oscillation amplitude exhibits a clear temperature dependence.
The temperature-dependent amplitude can be fitted by $\Delta T_C=\chi (t)\frac{\xi_0}{4r}T_C$, using a single $\xi_0=7.0$~nm.
This value is consistent with the obtained $\xi_0=7.3$~nm from the Nb ring with $r=300$~nm discussed in the main text.
As shown in Fig.~S4, the Little-Parks oscillations are well reproduced by Eq.~4 using the same value of $\xi_0$.
The consistent results obtained from Nb rings with different radii support the applicability of TAPS in $\Delta T_C$ calculation.

\begin{figure}
\centering
\includegraphics[width=16cm]{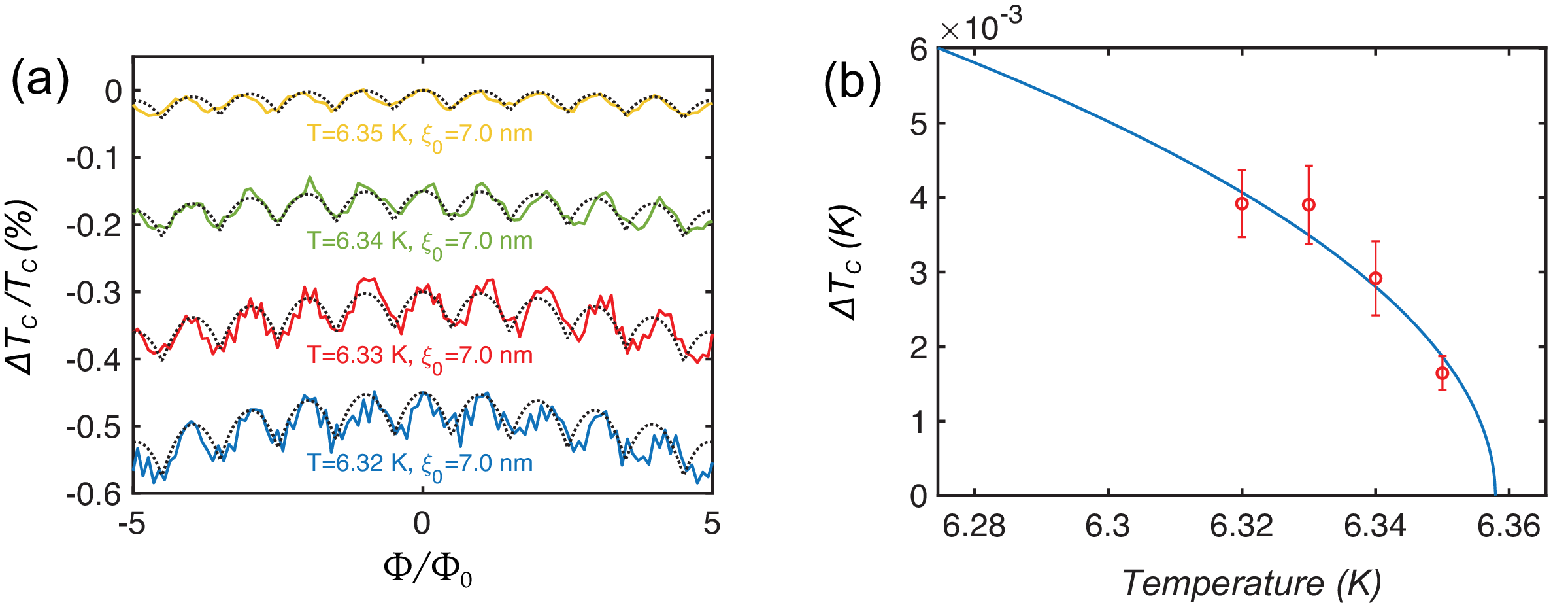}
\caption{(a) The Little-Parks oscillation curves of the Nb ring with $r=450$~nm at different temperatures. The solid lines are the experimental data and black dashed lines are fitting lines using Eq.~4. (b) Temperature-dependent oscillation amplitudes at different temperatures. The blue line is the fitting line by $\Delta T_C=\chi (t)\frac{\xi_0}{4r}T_C$ with $\xi_0=7.2$~nm.}
\end{figure}

Little-Parks measurements on $\beta$-Bi$_2$Pd rings with $r=450$~nm and $r=400$~nm also show the temperature-dependent oscillation amplitudes, as shown in Fig.S5.
The amplitudes can again be fitted by $\Delta T_C=\chi (t)\frac{\xi_0}{4r}T_C$ using a single coherence length, $\xi_0=19.0$~nm.
As an unconventional superconductor, $\beta$-Bi$_2$Pd exhibits an additional linear-in-field contribution to the Little-Parks background curve.
Therefore, the Little-Parks oscillation includes an additional term, which can be written as

 \begin{equation}
 \frac{\Delta T_c}{T_{c0}} = \chi(t) \left(\frac{\xi_0}{r} \right) \left[-\left(\frac{w}{2r} \right)^2 \left(\frac{\Phi}{\Phi_0} \right)^2+\frac{\rho_T}{\rho_L} \left(\frac{|\Phi|}{\Phi_0}\right) - \left(n-\frac{\Phi}{\Phi_0}\right)^2 \right].
 \label{eq:deltaTc2}
 \end{equation}

The fitting curves calculated using Eq.~S17 reproduce the experimental results shown in Fig.~2 and Fig.~S5.

\begin{figure}[htp]
\centering
    \includegraphics[width=16cm]{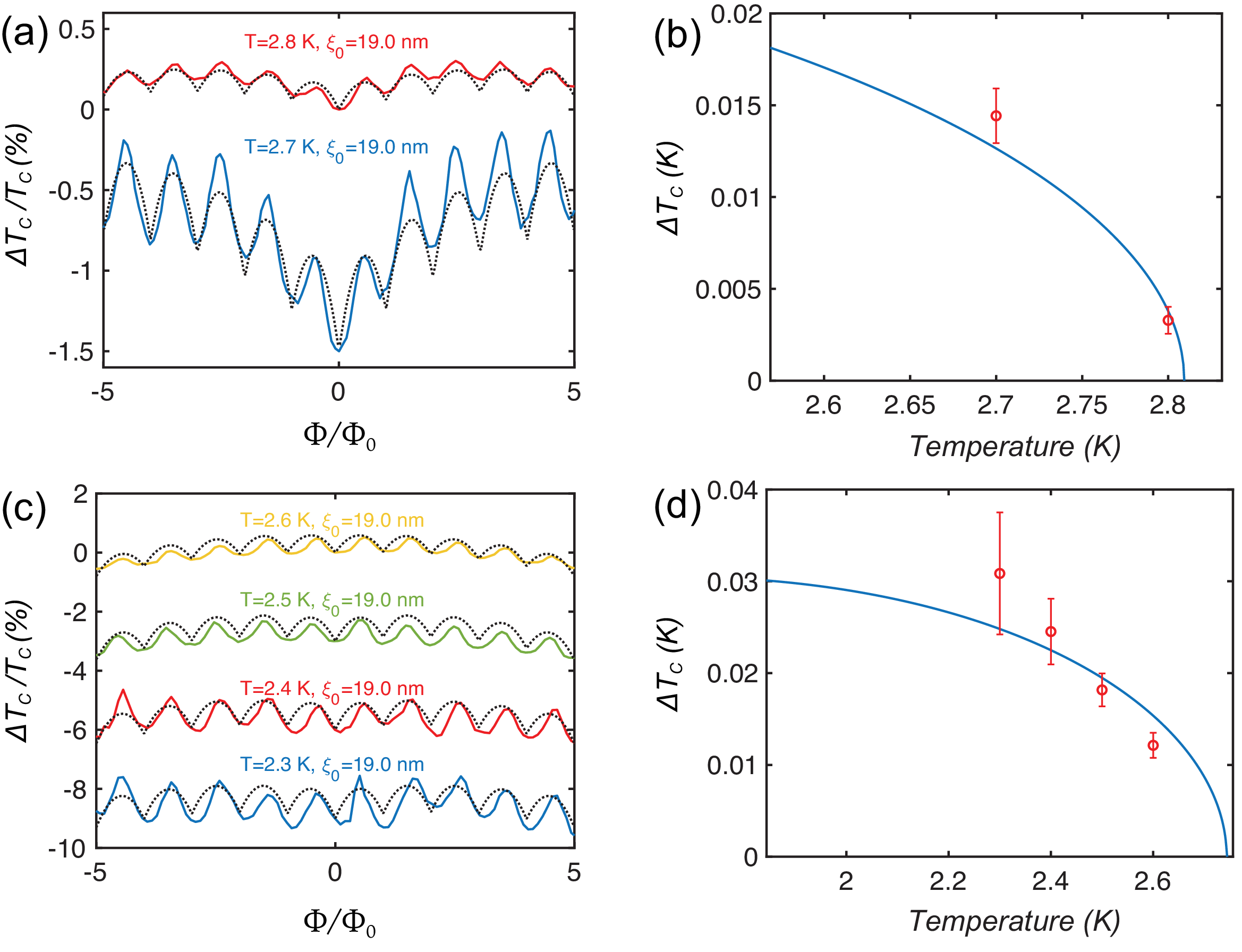}
\caption{The Little-Parks oscillation curves of the $\beta$-Bi$_2$Pd rings with $r=450$~nm (a) and $r=400$~nm (c) at different temperatures. The solid lines are the experimental data and the black dashed lines are fitting lines using Eq.~S17. Temperature-dependent oscillation amplitudes at different temperatures for (b) $r=450$~nm ring and (d) $r=400$~nm ring. The blue line is the fitting line by $\Delta T_C=\chi (t)\frac{\xi_0}{4r}T_C$ with a uniform coherence length $\xi_0=19.0$~nm. }
\end{figure}

Figure~S6 shows another $\alpha$-BiPd ring with $r=250$~nm. 
The Little-Parks oscillations again exhibit a clear temperature dependence, with a smaller amplitude at higher temperature. 
The temperature dependence of the oscillation amplitude can be well fitted by $\Delta T_C=\chi(t)\frac{\xi_0}{4r}T_C$, yielding $\xi_0\approx 13.5$~nm, in close agreement with the value obtained from the 400~nm $\alpha$-BiPd ring.

Although $\alpha$-BiPd is a non-centrosymmetric superconductor with a mixed spin-singlet and triplet pairing state, the Little-Parks oscillation curves show a simple parabolic background.
All Little-Parks oscillations in the $\alpha$ -BiPd rings can therefore be fitted by Eq.~4.

\begin{figure}[htp]
\centering
\includegraphics[width=16cm]{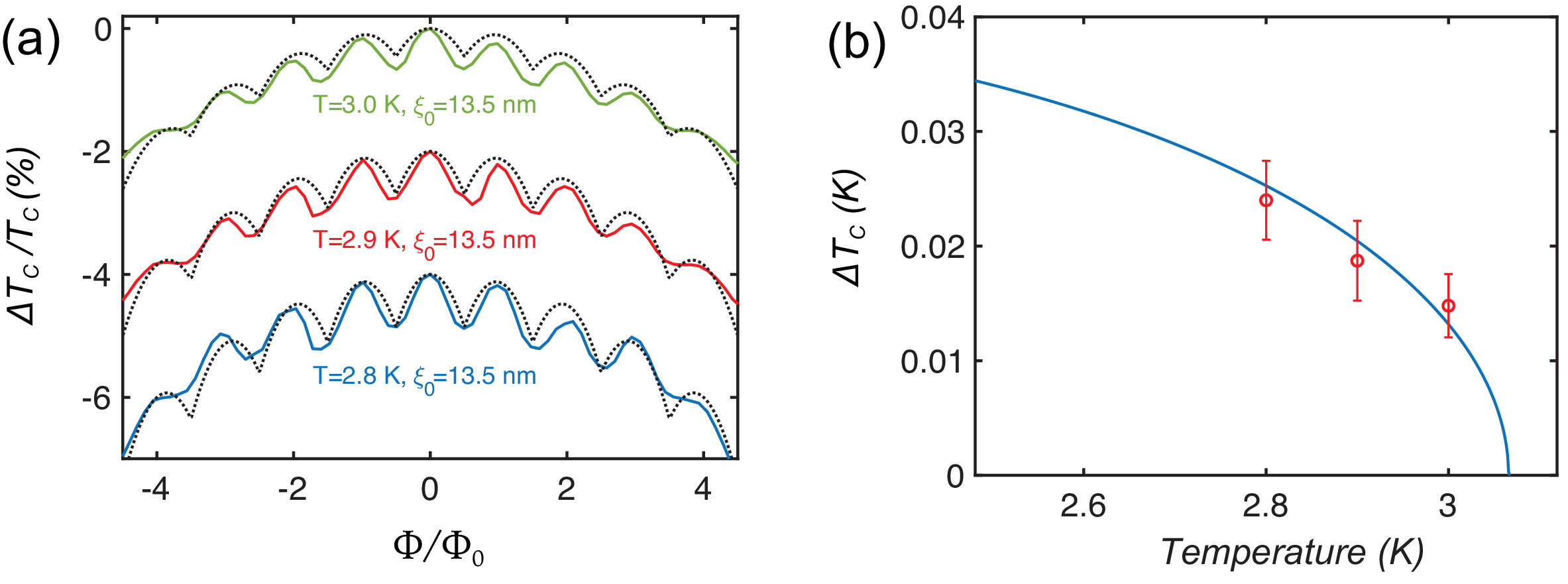}
\caption{(a) The Little-Parks oscillation curves of the $\alpha$-BiPd rings with $r=250$~nm at different temperatures. The solid lines are the experimental data and black dashed lines are fitting lines using Eq.~4. (b) Temperature-dependent oscillation amplitudes at different temperatures. The blue line is the fitting line by $\Delta T_C=\chi (t)\frac{\xi_0}{4r}T_C$ with $\xi_0=13.5$~nm, respectively. }
\end{figure}

Together, these measurements demonstrate that the temperature dependence of the Little-Parks amplitude is not specific to a particular material, ring radius, or background structure.
For both conventional and unconventional superconductors, the observed amplitudes can be consistently described by $\Delta T_C=\chi (t)\frac{\xi_0}{4r}T_C$.

\section{Analysis of the Little-Parks oscillations in high-$T_C$ superconductors using Eq.4}

The Little-Parks amplitude of La$_{1.84}$Sr$_{0.16}$CuO$_4$ can be fitted by $\Delta T_C=\chi(t)(\xi_0/4r)T_C$ with $\xi_0=1.1$~nm, as shown in Fig.~S7.
Using the same coherence length, the Little-Parks oscillation curves copied from the literature can be reproduced by Eq.~4.
This is in contrast to previous analyzes, where the oscillation curves could not be fitted by the conventional Eq.~1.

We also analyze the Little-Parks oscillations in YBa$_2$Cu$_3$O$_{7-\delta}$, shown in Fig.~S8 using data copied from a previous report.
The oscillation amplitude again exhibits a clear temperature dependence.
Fitting the amplitude yields a single coherence length of $\xi_0=3.2$~nm.
With this value, Eq.~4 reproduces the Little-Parks oscillation curves measured at different temperatures.

\begin{figure}
\centering
\includegraphics[width=16cm]{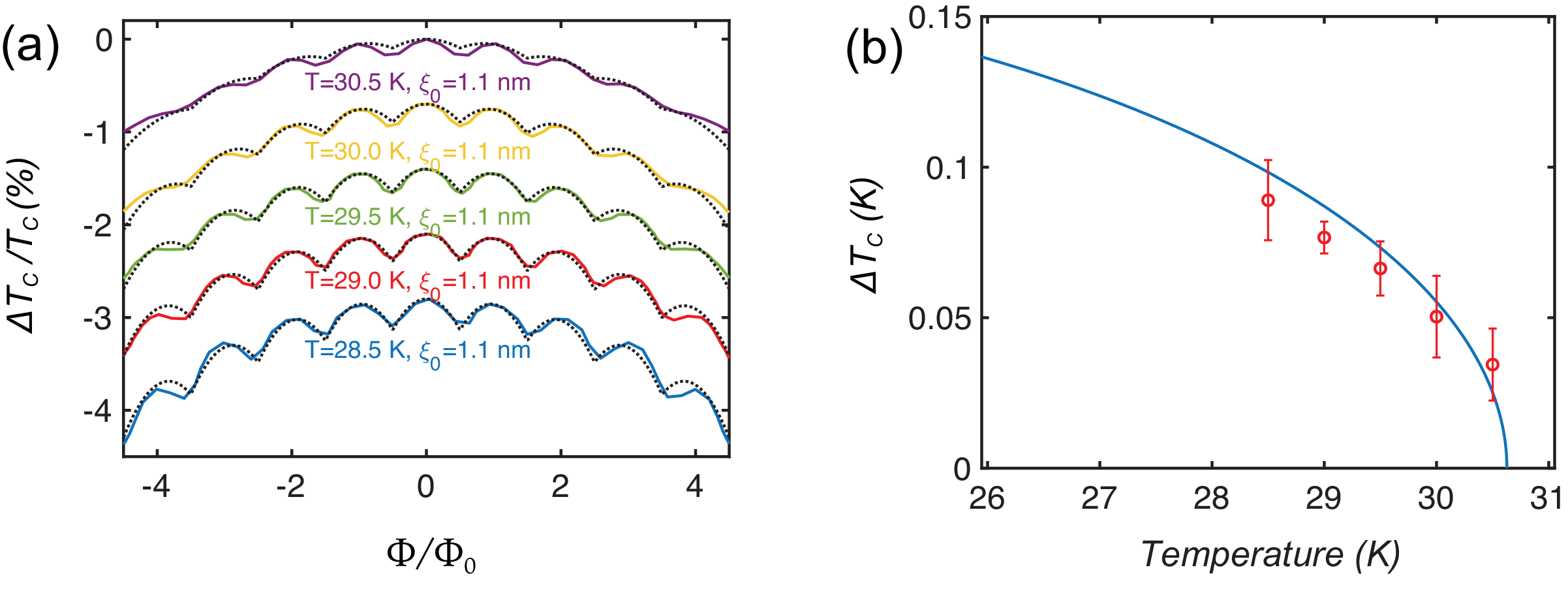}
\caption{(a) The Little-Parks oscillation curves of the La$_{1.84}$Sr$_{0.16}$CuO$_4$ ring copied from Ref.4. The solid lines are the experimental data and the black dashed lines are fitting lines using Eq.~4. (b) Temperature-dependent oscillation amplitudes at different temperatures. The blue line is the fitting line by $\Delta T_C=\chi (t)\frac{\xi_0}{4r}T_C$ with $\xi_0=1.1$~nm. }
\end{figure}

\begin{figure}
\centering
\includegraphics[width=16cm]{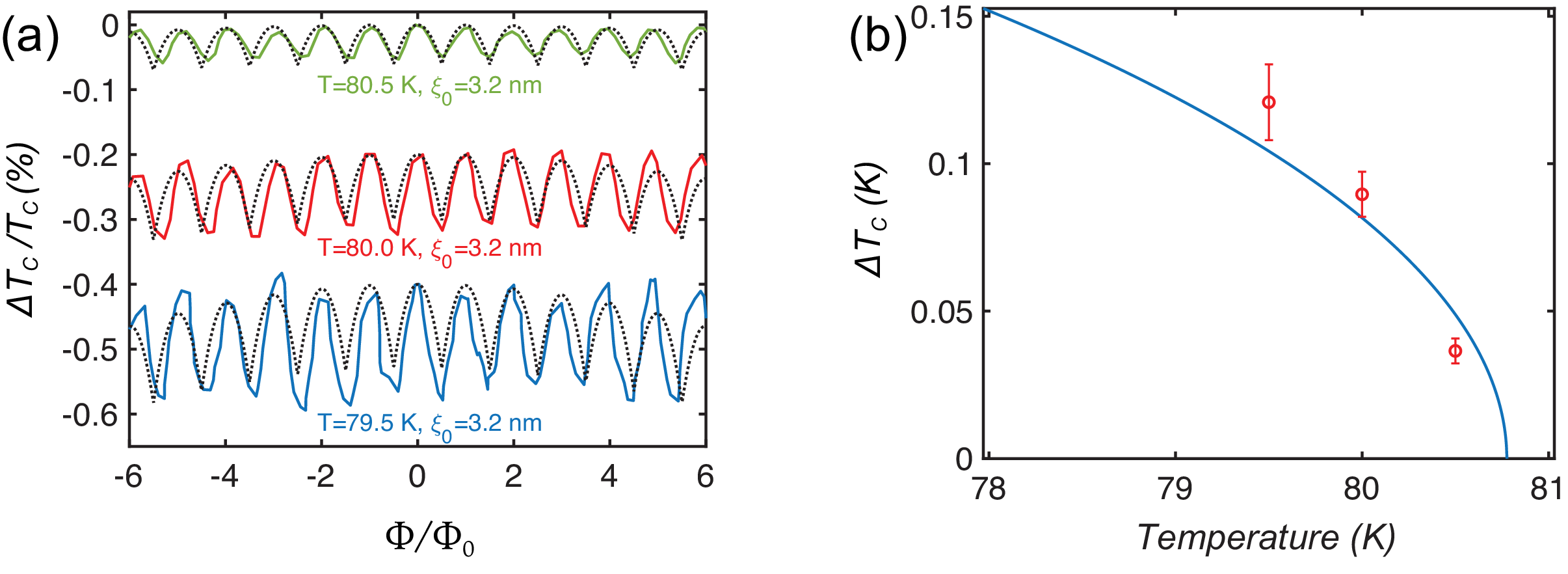}
\caption{(a) The Little-Parks oscillation curves of the YBa$_2$Cu$_3$O$_{7-\delta}$ ring copied from Ref.6. The solid lines are the experimental data and the black dashed lines are fitting lines using Eq.~4. (b) Temperature-dependent oscillation amplitudes at different temperatures. The blue line is the fitting line by $\Delta T_C=\chi (t)\frac{\xi_0}{4r}T_C$ with $\xi_0=3.2$~nm.}
\end{figure}

\endgroup


\clearpage

\end{document}